\documentclass[final,authoryear,5p,times,twocolumn]{elsarticle}
\usepackage{amssymb}
\usepackage{gensymb}
\usepackage[pdftex,bookmarks,bookmarksopen,colorlinks]{hyperref}
\usepackage{color,soul}
\usepackage{upquote}
\usepackage{enumitem}
\usepackage{etoolbox}
\usepackage{lineno}
\usepackage{graphics}
\usepackage{graphicx}
\usepackage{amsmath}
\usepackage{algorithmicx}
\usepackage{algpseudocode}
\usepackage{afterpage}
\usepackage{url}
\usepackage{tablefootnote}
\usepackage{epstopdf}
\journal{Astronomy and Computing}

\begin{document}

\begin{frontmatter}

\title{The impact of JPEG2000 lossy compression on the scientific quality of radio astronomy imagery}
  
\author{Sean~M.~Peters}
\ead{sean.peters.au@gmail.com}
\author{Vyacheslav~V.~Kitaeff\corref{cor1}}
\ead{slava.kitaeff@uwa.edu.au}
\cortext[cor1]{Corresponding author}
\address{International Centre for Radio Astronomy Research,\\
The University of Western Australia, \\
M468, 35 Stirling Hwy, Crawley 6009, WA, Australia}

\begin{abstract}
The sheer volume of data anticipated to be captured by future radio telescopes, such as, The 
Square Kilometer Array (SKA) and its precursors present new data challenges, including the
cost and technical feasibility of data transport and storage. Image and data
compression are going to be important techniques to reduce the data size. 
We provide a quantitative analysis of the 
effects of JPEG2000's lossy wavelet image compression algorithm on the
quality of the radio astronomy imagery data. This analysis is completed by 
evaluating the completeness, soundness and source parameterisation of the
\emph{Duchamp} source finder using compressed data. Here we
found the JPEG2000 image compression has the potential to denoise image cubes,
however this effect is only significant at high compression rates where the
accuracy of source parameterisation is decreased.
\end{abstract}


\end{frontmatter}
 
\section{Introduction} \label{cha:intro}

The upcoming Australian Square Kilometer Array Pathfinder (ASKAP) telescope~\citep{ASKAP..09} is anticipated
to capture spectral-imaging data-cubes (SIDCs)  several orders of magnitude larger than any
telescope ever before. The Square Kilometre Array (SKA)~\citep{2013arXiv1311.4288H} SIDCs will be even larger, 
in the order of tens of terabytes per hour of observations. 
The networking, storing, processing and interrogating of such large datasets poses new technical, 
financial, and management challenges.

Such large SIDCs cannot be processed or stored on local user
computers, even taking into account the projections for advances in HDD/SSD and network technologies,  the cost of data storage will be significant. Any means to reduce the data
volumes through compression will likely have significant benefits, especially in the reduction of the cost of the system.

\citealt{Kitaeff:2012:SCF:2286976.2286984} proposed the use of the JPEG2000 image 
standard as a method of storing SIDCs. JPEG2000 was first proposed in 
$1996$~\citep{JP2_proposal} with remote sensing and medical imaging being the most demanding imaging fields in mind. 
The JPEG2000 group, established in $1999$, has developed a 
standard containing 12 parts and defining the mechanisms not only for compression, but also for servicing 
and interrogating a broad range of imagery data\footnote{http://www.jpeg.org/jpeg2000/}. 

\citealt{2014arXiv1403.2801K} details the relevant features and mechanics of JPEG2000.

The most distinctive difference between JPEG2000 and the original JPEG image standard is the use of
the Discrete Wavelet Transform (DWT). The DWT is an algorithm to 
convert a signal to the time-frequency domain. Data in this domain lends itself
particularly well to quantisation and compression~\citep{bk:JP2}. As a result the JPEG2000 image standard 
can provide superior compression rates,
with much less loss in visual quality, than any other standardised image format
available. Additionally, JPEG2000 includes a variety of features especially
useful for extremely large images such as: progressive transmission, the
ability to decode any part of the image without having to decode the entire
image, adaptive encoding with different fidelities through a precinct mechanism, multiple resolutions from a single
master file. These features can significantly improve and enrich the interrogation of radio astronomy imagery data, as well
as, reduce the overall volume of the data.

While JPEG2000 may retain visual content effectively at high compression rates, the
effect of the lossy compression algorithm on the scientific quality of spectral image cubes
captured in radio astronomy needs to be understood. One important way to study the impact of lossy compression
on the quality of data is to see how well source finding algorithms identify the sources in compressed
SIDCs.

While the volumes of data are large, the information density of the data is rather low. 
For example, for studies using HI emission of extragalactic objects, the majority of sources of such an emission 
will appear in data occupying only a few pixels. Such sources will be rather sparsely populating the volume of data-cube.
The rest of imaging data is considered as noise.
To distil the information from the volume of data it is therefore necessary to identify and parameterise the sources. 

The development of fully 
automated source identification algorithms has recently become an intensive field of research~\citep{MNR:MNR20548, selavy, 2d1d,
CHNI, gamma, s+c}.
These algorithms would normally construct a catalogue of sources from an astronomical image.
Each entry in the constructed catalogue represents a source identified in the
image, and its determined parameters. The unknown systematic errors introduced into the catalogue 
at any stage are of course very problematic. Thus, the source finding algorithms themselves need to be 
investigated, as well as, any additional data processing such as compression. The completeness and soundness 
are the main measures of how successful the algorithm is, in finding the sources~\citep{art:SF-comparison}. Such a set of metrics could be usefully utilised 
to study the impact of lossy compression on the quality of data, providing the guidance to the developers
of the SKA as to whether JPEG2000's lossy compression may be safely used.

The rest of the paper is laid out as follows. Section~\ref{cha:exp} details the dataset used and methodology taken in 
conducting our
experiment. Section~\ref{cha:res} provides the results and detailed discussion
on the experiments performed in this paper. Finally, we draw some conclusions in Section~\ref{cha:conc}.

\section{Methodology} \label{cha:exp}

\subsection{Synthetic dataset} \label{d:sim}

In order to test the impact of compression on the scientific quality of data, we have chosen 
to use a simulated SIDC in preference to real radio astronomy imagery. By using 
a simulated set of galaxies to create our imagery the true catalogue of
sources within the image cube is known with a higher certainty. This helps with
measuring, not only the accuracy of source identification after image compression,
but source parameterisation as well. The Deep Investigations of Neutral Gas Origins (DINGO)~\citep{2009pra..confE..15M}
synthetic SIDC was used in all our tests.

DINGO is one of the planned surveys to be performed by the 
ASKAP telescope. The synthetic SIDC was generated using an analytical galaxy 
simulation as described by~\citet{duffy}, as part of survey planning. Significant effort was 
placed in creating a plausible DINGO survey simulation, as
well as correctly inserting the sources while simulating instrumental noise and
errors. A brief summary of the steps is outlined below:
\begin{enumerate}
\item An analytical model for each of the different types of galaxies was
produced. This was followed by a model of the distribution of these galaxies
throughout the DINGO survey space.
\item A cosmological simulation was produced using the analytical models of
galaxies and their distribution. This provided position, flux, HI content etc. 
for almost 4 million galaxies, the vast majority of which would be 
unidentifiable. These sources formed a true catalogue in our experiments. Many of the 
simulated sources, however, are too faint to be observable with ASKAP. For the experiments performed 
in this paper a filter was applied on the true source catalogue ~\citep{duffy}.
\item The galaxies were then injected into an empty image cube. At this point the 
dataset contains a perfectly clean set of galaxies distributed throughout the cube.
\item Mock visibility data (radio telescopes capture data in the visibility
domain - the Inverse Fourier Transform of the image cube) was then generated 
from the image cube using the \emph{Miriad} software~\citep{miriad}.
\item The visibility data was then convolved with the dirty ASKAP beam, to
introduce instrumental noise.
\item The resultant visibility data was passed through the Fourier Transform
and then convolved with the clean image cube.
\item Finally Gaussian noise was distributed over the entire image cube following
the profile of thermal noise expected in the ASKAP telescope.
\end{enumerate}

The entire datacube is approximately 1 Terabyte in size with dimensions
$3,600\times3,600\times23,060$. The spatial region represented approximately
60deg$^2$. Each frequency channel in the cube (third dimension) 
has a width of 18.518 KHz. 

\emph{Duchamp}~\citep{MNR:MNR20548} and several other used tools that were tested, required 
too much memory and computational power to
feasibly process the entire cube at once, while meaningfully exploring the required parameter space of 
JPEG2000. Therefore extracts were taken from the original cube and each was tested independently. 

The lower the frequency, the farther away the galaxies are in the cube due to 
the cosmological redshift of HI, and therefore the sources appear fainter towards 
the low frequency end of the cube.  It was therefore important to test several subsets of data at 
different frequency ranges.

A subcube should also contain a sufficiently large number of sources in order to provide 
statistical significance in the tests. It was decided, arbitrarily, that the source finder should 
be able to identify at least 50 sources within the subcube.

At the low frequency end of DINGO datacube the signal-to-noise ratio for the sources becomes very small. 
It was established in tests that \emph{Duchamp} was able to identify very few sources
above the frequency plane $\sim$15000, which was selected as a lower boundary for the frequency axes.

Table~\ref{tab:subcubes} shows the three subcubes extracted from DINGO cube to 
perform the tests. The $Z$ references the frequency axes of the dataset.
Larger $Z$ corresponds to the lower frequency.

Table~\ref{tab:subcubes_stats} shows the mean and variance of the datasets A, B and C.

\begin{figure}
  \centering
  \includegraphics[width=90mm]{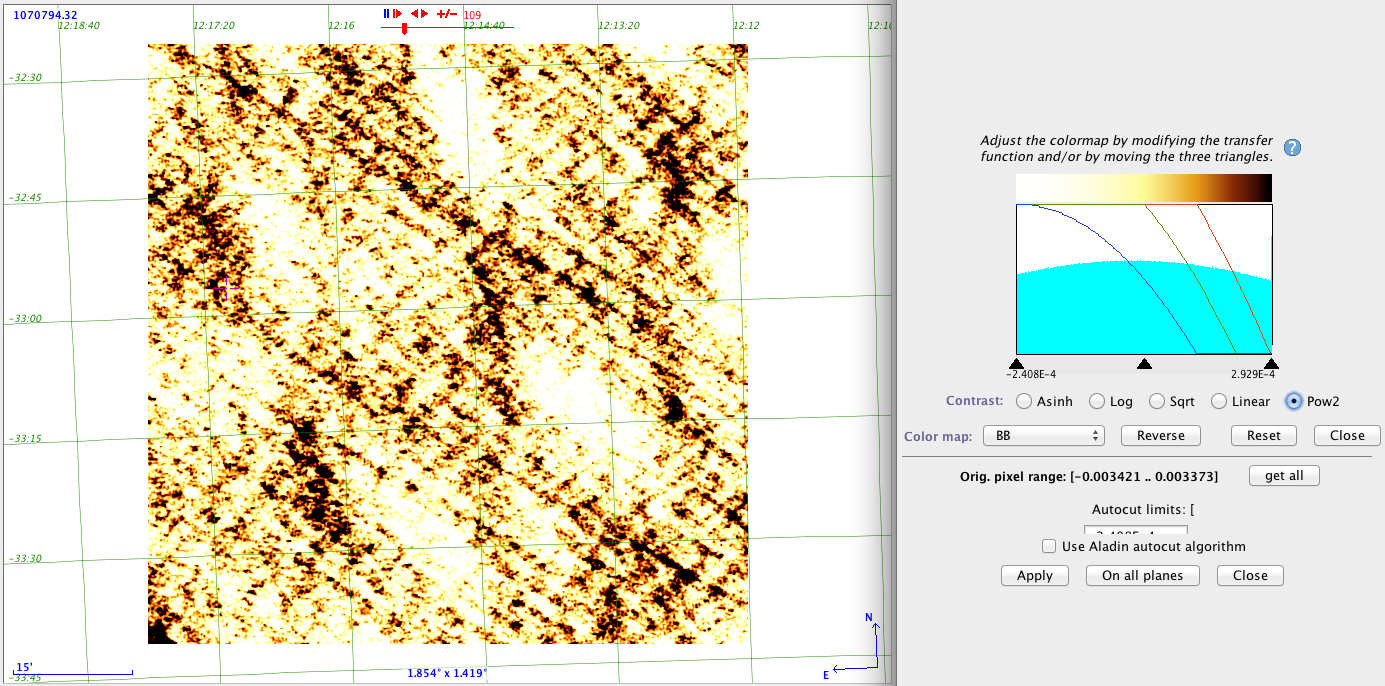}
  \caption{Typical single frequency frame of synthetic DINGO cube with a few hundred sources (darker pixels). 
The colours are artificial.}
  \label{fig:Plane-image}
\end{figure}

\begin{figure}
  \centering
  \includegraphics[width=90mm]{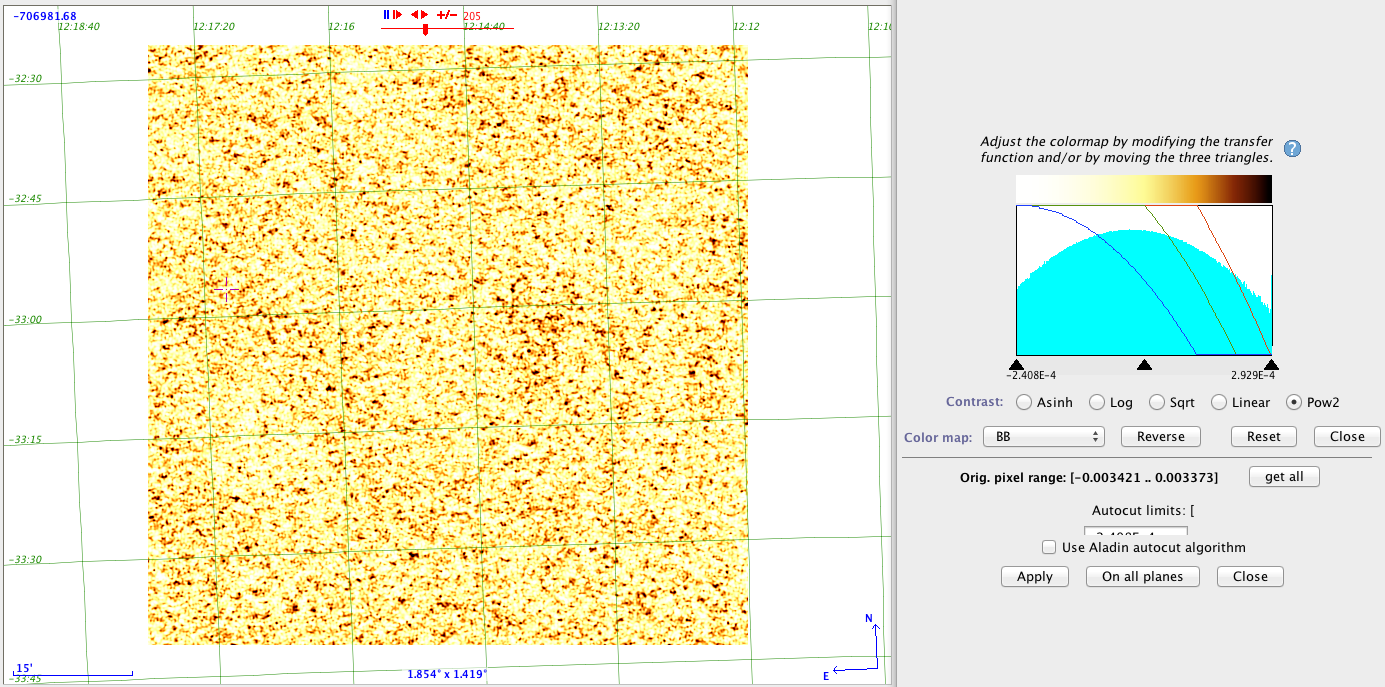}
  \caption{Corrupted by the instrumental effects single frequency frame of synthetic DINGO cube. 
The colours are artificial.}
  \label{fig:Plane-image-corrupted}
\end{figure}

Figures~\ref{fig:Plane-image} and \ref{fig:Plane-image-corrupted} depict a typical and 
corrupted by the instrumental effects frames of the used synthetic cube containing
hundreds of point sources (dark pixels). For each of such sources HI spectral line 
profile exist in the $Z$ dimension.

\begin{table}[ht]
\centering
\begin{tabular} {|l|l|l|l|l|l|l|}
\hline
Dataset & $X_0$ & $Y_0$ & $Z_0$ & $\Delta X$ & $\Delta Y$ & $\Delta Z$ \\
\hline
A & 0 & 0 & 4000 & 3600 & 3600 & 100 \\
B & 0 & 0 & 7000 & 3600 & 3600 & 100 \\
C & 0 & 0 & 10100 & 3600 & 3600 & 100 \\
\hline
\end{tabular}
\caption{Subcubes selection within the DINGO datacube}
\label{tab:subcubes}
\end{table}

\begin{table}[ht]
\centering
\begin{tabular} {|l|l|l|}
\hline
Dataset & Mean & Variance ($\times10^{-5}$) \\
\hline
A & $-8.534\times10^{-10}$ & 3.650 \\
B & $9.588\times10^{-11}$ & 3.579 \\
C & $6.252\times10^{-09}$ & 3.532 \\
\hline
\end{tabular}
\caption{Mean and Variance of each dataset.}
\label{tab:subcubes_stats}
\end{table}

\subsection{JPEG2000 encoding/decoding} \label{d:enc}

\begin{figure}
  \centering
  \includegraphics[width=90mm]{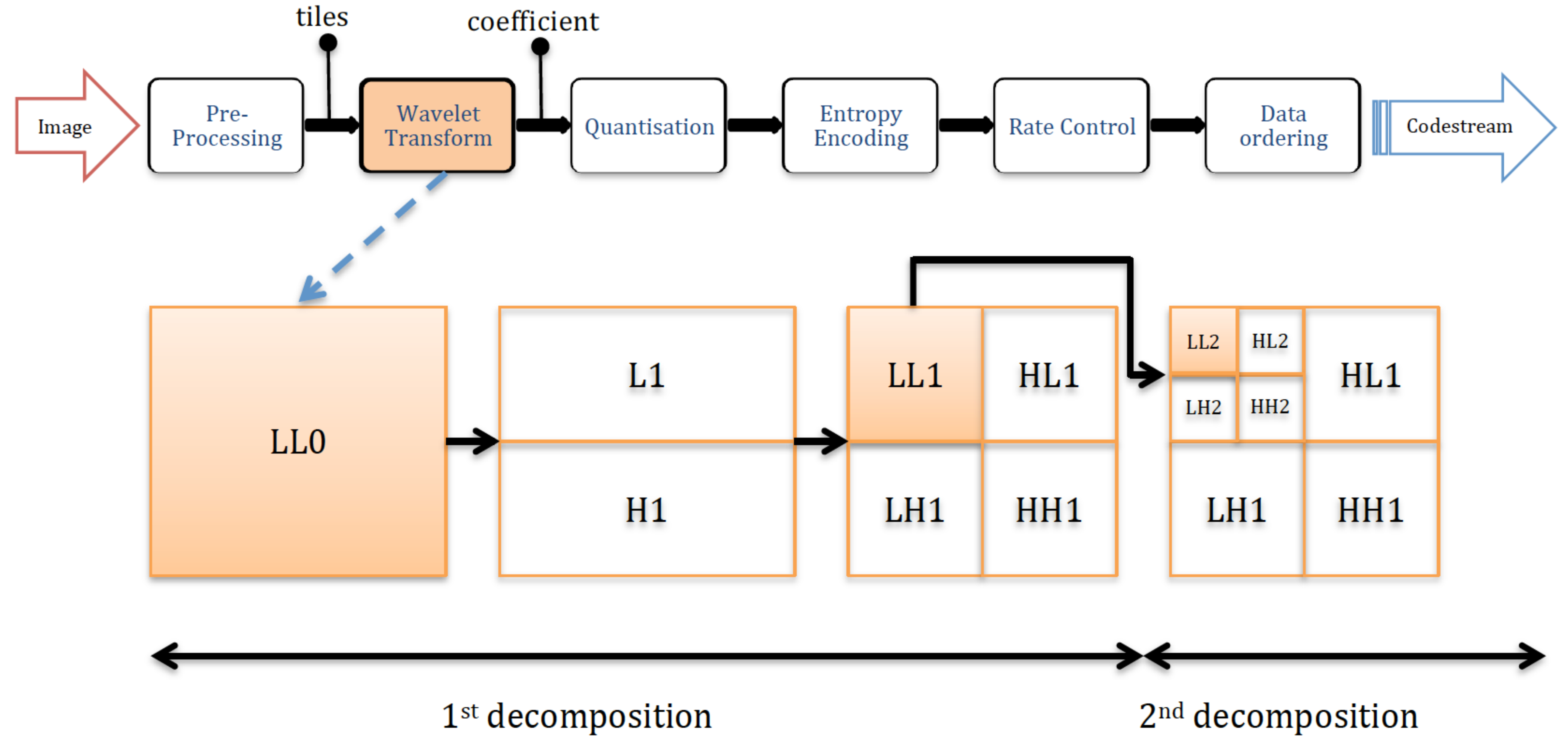}
  \caption{JPEG2000 encoding is based on discrete wavelet transformation, scalar quantisation,
context modelling, arithmetic coding and post-compression rate allocation.}
  \label{fig:jpeg2000-encoding}
\end{figure}

Unlike the binary compression available through \texttt{cfitsio} or HDF5, JPEG2000 is a true image
compression that takes advantage of the multidimensionality of data. Figure~\ref{fig:jpeg2000-encoding} depicts 
the stages of encoding in JPEG2000.

In the first stage, pre-processing is performed. Pre-processing actually contains three substages: 
Tiling, Level Offset, Reversible/Irreversible Color Transform. This stage prepares the data to correctly perform
the Wavelet Transform. During the Wavelet Transform, image components are passed recursively through the 
low pass and high pass Wavelet filters. This enables an intra-component decorrelation that
concentrates the image information in a small and very localised area. It enables the multi-resolution
image representation. The result is that 4 sub-bands with the upper left one $LL$ on Figure~\ref{fig:jpeg2000-encoding} containing
all low frequencies (low resolution image), $HL$ containing vertical high frequencies, LH containing horizontal high frequencies, and $HH$
containing diagonal high frequencies. Successive decompositions are applied on the low
frequencies $LL$ recursively as many times as desired.

By itself the Wavelet Transform does not compress the image data; it restructures the image information so
that it is easier to compress. Once the Discrete Wavelet Transform (DWT) has been
applied, the output is quantified in Quantisation unit.

Before coding is performed, the sub-bands of each tile are further partitioned into small code-blocks (e.g. 64x64
or 32$x$32 samples) such that code blocks from a sub-band have the same size. Code-blocks
are used to permit a flexible bit stream organisation.

The quantised data is then encoded in the Entropy Coding unit.
The Entropy Coding unit is composed of a Coefficient Bit Modeller and the Arithmetic Coder itself.
The Arithmetic Coder removes the redundancy in the encoding of the data. It assigns short code-words to
the more probable events and longer code-words to the less probable ones.
The Bit Modeller estimates the probability of each possible event at each point in the coding stream.

At the same time as embedded block coding is being performed, the resulting bit streams
for each code-block are organised into quality layers. A quality layer is a collection of some
consecutive bit-plane coding passes from all code-blocks in all sub-bands and all components, 
or simply stated, from each tile. Each code-block can contribute an arbitrary number
of bit-plane coding passes to a layer, but not all coding passes must be assigned to a quality
layer. Every additional layer successively increases the image quality.

Once the image has been compressed, the compressed blocks are passed over to the Rate Control unit 
that determines the extent to which each block's embedded bit stream should be truncated in order to achieve 
the target bit rate. The ideal truncation strategy is one that minimises distortion while still reaching the target bit-rate.

In Data Ordering unit, the compressed data from the bit-plane coding 
passes are first separated into packets. One packet is generated for each precinct in a tile. A precinct is
essentially a grouping of code blocks within a resolution level. Then, the packets are multiplexed 
together in an ordered manner to form one code-stream. There are five built-in ways to order the packets, called 
progressions, where position refers to the precinct number:
\begin{itemize}[leftmargin=*]
\item Quality: layer, resolution, component, position
\item Resolution 1: resolution, layer, component, position
\item Resolution 2: resolution, position, component, layer
\item Position: position, component, resolution, layer
\item Component: component, position, resolution, layer
\end{itemize}

The decoder basically performs the opposite operations of the encoder.

The details and mathematics of JPEG2000 encoding can be found in \citealt{Gray}, \citealt{Adams2001}, or \citealt{Li2003}.

\subsection{Software} \label{sec:tools}

As JPEG2000 has not yet seen use in the radio astronomy
domain many of the tools required, such as, source finders or image viewers did 
not provide direct support for the image standard. Several tools, therefore, were
developed to support the experiments. 

As the original dataset was stored using the HDF5 and FITS formats, 
we developed the \emph{skuareview-encoder} and \emph{skuareview-decoder} software\footnote{https: //github.com/ICRAR/SkuareView}, 
to encode/decode the data to and from JPEG2000 using the JPEG2000 KDU library from Kakadu Software\footnote{http://www.kakadusoftware.com/}.

The \emph{Duchamp} source finder was used as the source identification 
software. \emph{Duchamp} is the predecessor of the \emph{Selavy} source finder
(the source finder anticipated to be used in ASKAP and the SKA). Unfortunately
\emph{Selavy} was not made publicly available at the time of these works.

There are two particularly interesting aspects of \emph{Duchamp} that are
directly related to the experiments in this paper.

Firstly, \emph{Duchamp} has the option to perform a wavelet reconstruction on the
image cube before sources are searched for. This wavelet reconstruction aims to
denoise the image. The ``\emph{algorithme \`{a} trous}''~\citep{atrous} is the
wavelet transform used for this purpose as it maintains shift invariance
where other wavelet transforms do not.
The JPEG2000 image standard uses the Discrete Wavelet Transform (DWT) as a step
in its compression algorithm. The DWT does not maintain shift invariance, however
the algorithm is significantly less redundant as it includes
subsampling~\citep{DWT-shift_variance}.

Secondly, \emph{Duchamp} is known to be moderately inaccurate when parameterising
sources found in the image~\citep{duchamp-testing}. This must therefore be
considered when measuring the accuracy of source parameterisation after JPEG2000
image compression.

\subsection{Experiment design} \label{sec:exp11}

\subsubsection {Completeness and Soundness}

The experiment aimed to provide a measure of the effect of the JPEG2000 image
compression algorithm on the scientific quality of radio astronomy data. Here we
define the scientific quality as the usefulness of the data to astrophysicists and
their experiments. More specifically, we note that the only features important
for DINGO science, found within a raw
radio astronomy image cube are the HI sources within the cube; the vast majority
of which represent the HI emission of distant galaxies. Everything else
in the image cube is regarded as noise. As such, it is just the HI sources in the
image that
needed to be identified. This identification process needed to be as complete and
as sound as possible, and each identified object needed to be parameterised as
accurately as possible.

Source identification across an image cube is said to be complete if every
source in the trueset catalogue is included in the identified set. The equation for
completeness $C$ is given by

\begin{equation}
C = \frac{P_{true}}{N}
\label{eq:comp}
\end{equation}

where $P_{true}$ represents all true positive identifications, and $N$ is the number of
sources in the trueset catalogue.

The source identification process is sound if every element in the identified
set is included in the trueset catalogue. The equation for soundness $S$ is given by

\begin{equation}
S = \frac{P_{true}}{P_{total}}
\label{eq:sound}
\end{equation}

where $P_{total}$ is the total number of detections that includes both, positive and negative detections.

To obtain the effect the JPEG2000 compression has on the completeness and
soundness of the source finders we use measure ``completeness difference''
$\psi$, and ``soundness difference'' $\omega$. These measurements are defined 
by

\begin{gather}
\psi = C_{compressed} - C_{original} \\
\omega = S_{compressed} - S_{original} 
\label{eq:cs_diff}
\end{gather}

where $C_{compressed}$ and $S_{compressed}$ are the completeness and soundness respectively
calculated on the compressed dataset,
and $C_{original} $ and $S_{original} $ are the completeness and soundness respectively
calculated on the original dataset.

\subsubsection{Source Parameterisation}

After identification, a source also needs to be parameterised. Each source is
attributed to a location in galactic coordinates Right Ascension (RA),
Declination (Dec) and Frequency. Beyond
this a variety of parameters are used to specify the attributes of each source. 
In this test we investigated the parameters listed below: 

\begin{itemize}
\item Right Ascension
\item Declination
\item Frequency
\item Right Ascension Width
\item Declination Width
\item Frequency Width 
\item Integrated Flux 
\label{src_params}
\end{itemize}

We measured the difference in the performance of source parameterisation between
the compressed dataset and the control dataset $\gamma$, for some JPEG2000 
parameter $k$ and some source parameter type $p$ using the following equation

\begin{equation}
\gamma _k = \text{RMSE}_o(p) - \text{RMSE}_k(p) \\
\label{eq:param_diff}
\end{equation}

where RMSE$_o$ is the Root Mean Square Error (RMSE) between the source parameter value 
identified with the original dataset, and the true source parameter, while 
RMSE$_k$ is the RMSE between the source parameter value identified with the dataset
compressed with JPEG2000 parameter $k$, and the true source parameter value. 

The value $\gamma _k$ will thus be positive if the compressed dataset allows for more
accurate parameterisation and negative otherwise.

For example, to measure the effect of JPEG2000's image compression 
algorithm on the parameterisation
of sources with respect to the frequency width parameter of all sources
identified, we calculate the RMSE of the frequency 
width of each source identified within the dataset by \emph{Duchamp} against the trueset
catalogue's frequency width for the respective parameter. We then perform the
same RMSE calculation using the sources identified by \emph{Duchamp} after compression.
If the difference between the resultant RMSE from the compressed dataset, and the
RMSE from the original dataset is positive, then the compressed dataset has
allowed \emph{Duchamp} to provide more accurate parameterisation. However, if this
difference is negative this would imply the compression has damaged the
scientific quality of the data.

\subsubsection{Cross matching} \label{ssec:cross}

In order to correctly measure source parameterisation and the accuracy of the
source finder, we need to ensure that the sources retrieved by the
\emph{Duchamp} are attributed to the correct source in the trueset catalogue. 

This was done by iterating over all pairs $(u_i,v_j)$ where $u_i$ is the $i^{th}$
source obtained from the \emph{Duchamp} source finder and $v_j$ is the $j^{th}$ source
from the true catalogue. A pair was considered a match if the following conditions were
true:

\begin{enumerate}
\item Condition 1
  \begin{enumerate}
    \item The center of $u_i$ was contained within the bounds of $v_j$, 
    \item \textbf{OR} the center of $u_i$ was within 3 voxels of the center of $v_j$,
  \end{enumerate}
\item Condition 2
  \begin{enumerate}
    \item if multiple sources from the \emph{Duchamp}
    catalogue were potential matches with a single $v_j$, the $u_i$ with the closest
    center to $v_j$ was chosen.
  \end{enumerate}
\end{enumerate}

\subsubsection{Comparison of JPEG2000 against the Wavelet
Reconstruction in Duchamp}

This experiment was intended to evaluate the JPEG2000 lossy compression as a
denoising tool, by drawing a comparison between JPEG2000 lossy compression and 
the wavelet reconstruction algorithm used in \emph{Duchamp}.

The differences in completeness, soundness and source parameterisation were
calculated between \emph{Duchamp} with a wavelet reconstruction and
\emph{Duchamp} without the wavelet reconstruction. These results were then graphed
alongside the differences in completeness, soundness and source parameterisation
of \emph{Duchamp} between compressed imagery and uncompressed imagery.

Only one set of parameters was used for the wavelet reconstruction
as described in Table~\ref{tab:src_params} due to the fact that as the \emph{Duchamp} wavelet 
reconstruction was computationally exhaustive. These parameters were
chosen with reference to experiments performed on similar
datasets~\citep{duchamp-testing,art:SF-comparison}. 

\begin{table}[ht]
\centering
\begin{tabular} {|p{0.27\columnwidth}|p{0.09\columnwidth}|p{0.54\columnwidth}|}
\hline
Parameter Name & Value & Comment \\
\hline
minVoxels & 7 & Minimum voxels required to identify a source\\
flagAdjacent & true & Identified objects are merged using adjacency\\
snrCut & 5 & Threshold in multiples of standard deviation for the Isotropic Undecimated Wavelet Transform (IUWT)~\citep{IUWT} \\
scaleMin & 2 & Minimum wavelet scale in the IUWT~\citep{IUWT}\\
\hline
\end{tabular}
\caption{Duchamp source finder parameters}
\label{tab:src_params}
\end{table}

\subsubsection{JPEG2000 parameter space}

In our experiment we selected four of the most important parameters to investigate from 
the JPEG2000 parameter space:

\begin{itemize}
\item The \emph{Quantisation step size} is extremely influential on the
compression ratio and lossiness of the JPEG2000 compression algorithm. Smaller
step sizes will result in less quantisation and therefore less lossy
compression. By exploring this parameter we can observe the influence of the
JPEG2000 compression algorithm at different levels of lossiness.
\item The number of \emph{levels in the tree of the DWT} influences the 
structure of the wavelet domain before quantisation and compression.
\item \emph{Precincts} partition the image cube into rectangles that are each
encoded independently. This will effect how the wavelet domain will be supplied
to quantisation and compression resulting in differing compression ratios.
\item The \emph{Code block size} effects the size of the most granular partition
in the JPEG2000 compression algorithm. Large block sizes will provide more
opportunity for compression.
\end{itemize}

These parameters were explored as described in Table~\ref{tab:jp2_params}.

\begin{table}[htbp]
\centering
\begin{tabular}
{|l|l|l|l|l|l|}
\hline
Parameter Name & Default & Start & End & Step \\
\hline
Quantization step size & $1/256$ & $10^{-6}$ & $0.01$ & $\times \sqrt[4]{10}$ \\ 
\hline
DWT levels & 5 & $1$ & $32$ & $+4$ \\
\hline
Precincts & $2^{15}$ & $64$ & $1024$ & $\times 2$ \\
\hline
Code block size & 64 & $4$ & $64$ & $\times 2$ \\
\hline
\end{tabular}
\caption{JPEG2000 compression parameters iterated over in our experiment.}
\label{tab:jp2_params}
\end{table}

\subsubsection{Procedure}

A script was developed in order to perform this experiment over multiple JPEG2000
parameters and 
multiple datasets. The process is described below using the following function
definitions;

\begin{itemize}
\item Duchamp($D$) takes dataset $D$ and returns catalogue $C$.
\item Process($C$) takes catalogue $C$ and retrieves the completeness, soundness
and source parameterisation results $R$. 
\item Encode takes dataset $D$, parameter type $j$ and parameter value $i$ and
returns the JPEG2000 lossily compressed image.
\item Decode takes the JPEG2000 lossily compressed image and decodes it into a
FITS (required by \emph{Duchamp}) dataset.
\end{itemize}

\begin{algorithmic}
  \State $C_t \gets \text{true catalogue}$
  \State $D_o \gets \text{original dataset}$
  \State $C_o \gets \text{Duchamp}(D_o)$
  \State $R_o \gets \text{Process}(C_o)$
  \ForAll{JPEG2000 parameter types $j$ in Table~\ref{tab:jp2_params}}
    \ForAll{values $i$ for parameter type $j$ in Table~\ref{tab:jp2_params}}
      \State $D_j^i \gets \text{Decode}(\text{Encode}(D_o,i,j))$
      \State item $C_j^i \gets \text{Duchamp}(D_j^i)$
      \State item $R_j^i \gets \text{Process}(C_j^i)$
    \EndFor
    \ForAll{$i$} \\
      \hspace{12mm}Graph($R_j^i - R_o$)
    \EndFor
  \EndFor
\label{code:exp1}
\end{algorithmic}

3D WDT encoding may improve the efficiency of encoding \citep{Delcourt:2011:EFB:2436496.2436501},
however, KDU library only enables 2D encoding with a possibility to use the image components 
as another dimension. Thus, in all of our tests frames of SIDC were encoded independently into the
image components using 2D WDT. 

\section{Results and Discussion} \label{cha:res}

The presented results are based in intensive testing and many runs with different
parameters. Due to the large data volumes many experiments require many hours
to complete.

\subsection{Compression Ratio and RMSE} \label{ssec:cr_rmse}

Fundamentally, our most important choice when using any lossy compression
algorithm, is to what degree the data can be compressed without the introduction of a significant error. 
We explored a variety of parameters in our experiments and 
measured the effect of a change on each parameter on the scientific quality of
our dataset. This effect on scientific quality can be more intuitively understood with reference to each
parameter's effect on compression ratio and RMSE. 

\begin{figure}[ht]
  \includegraphics[width=\columnwidth]{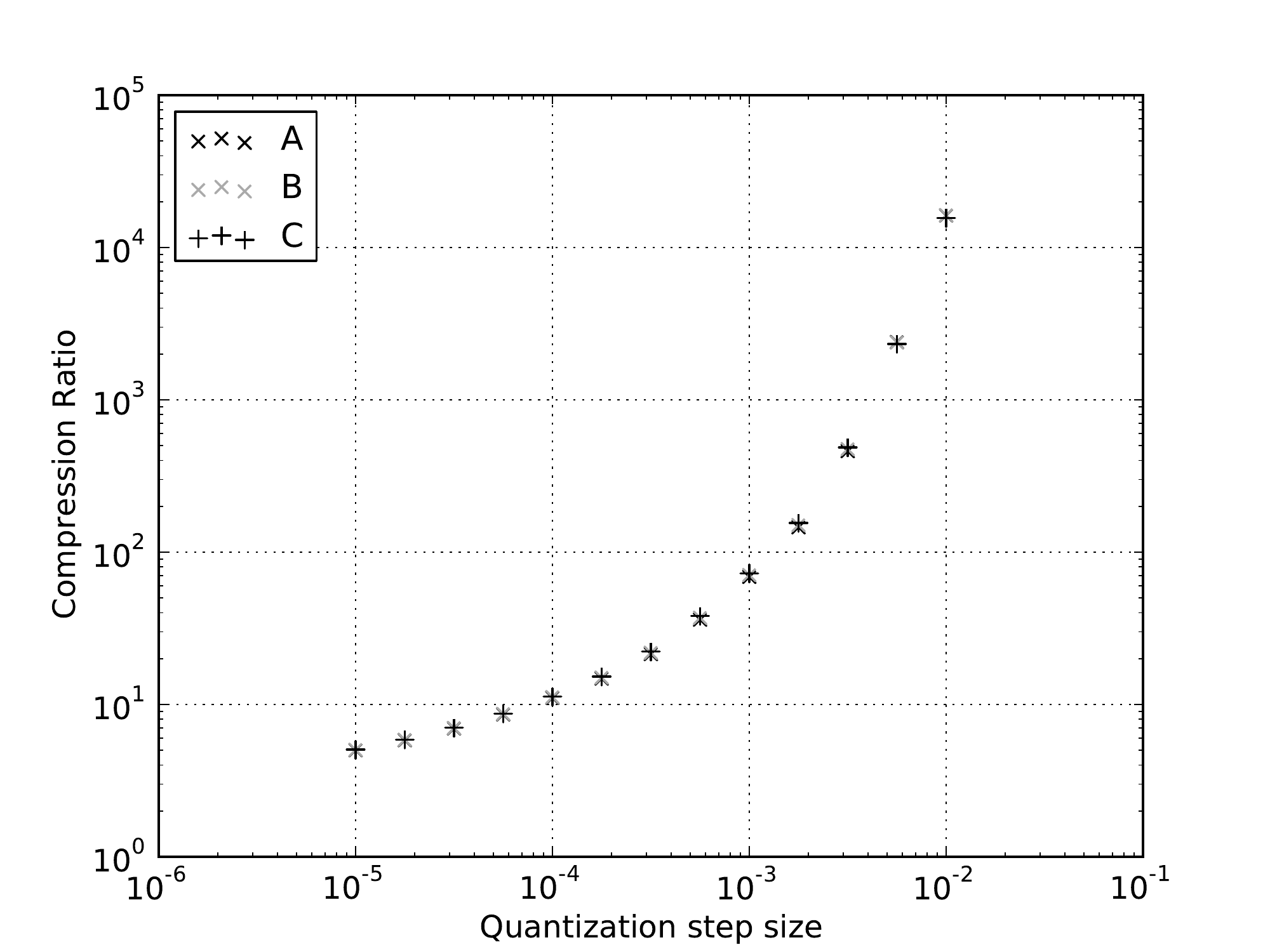}
  \caption{Compression Ratio vs Quantisation step size for datasets $A$, $B$
    and $C$.}
  \label{fig:qstep-compress}
\end{figure}

Figure~\ref{fig:qstep-compress} demonstrates a direct exponential correlation between 
the quantization step size and the compression ratio of the JPEG2000 compression
algorithm. This correlation was consistent across each dataset used. The
quantization step size was observed over our results to be the most
influential JPEG2000 parameter on the compression ratio and RMSE.

\begin{figure}[ht]
  \includegraphics[width=\columnwidth]{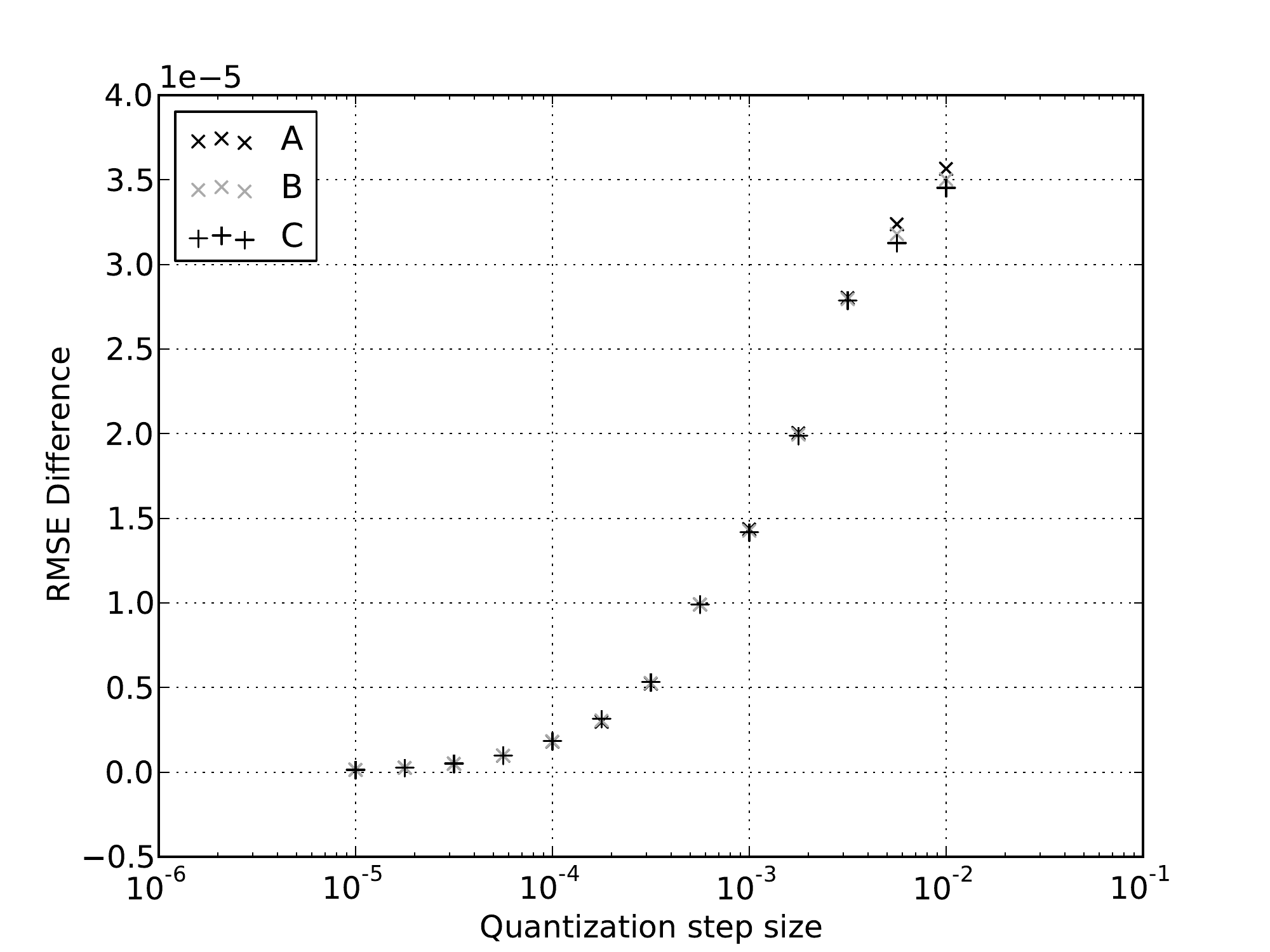}
  \caption{RMSE Difference vs Quantisation step size for datasets $A$, $B$ and $C$.}
  \label{fig:qstep-rmse}
\end{figure}

As it can be expected, the RMSE Difference, which is a measure
of difference of the original and compressed versions of the image, is increasing with
the quantisation step size (see Figure~\ref{fig:qstep-rmse}). This on its own
does not indicate the ability of a source finder to detect the sources. In fact, the increasing 
RMSE difference of a noisy image (Fig.\ref{fig:Plane-image}) indicates a noise filtering
effect produced by the compression. The question is how this impacts the sources in the image?

\begin{figure}[ht]
  \includegraphics[width=\columnwidth]{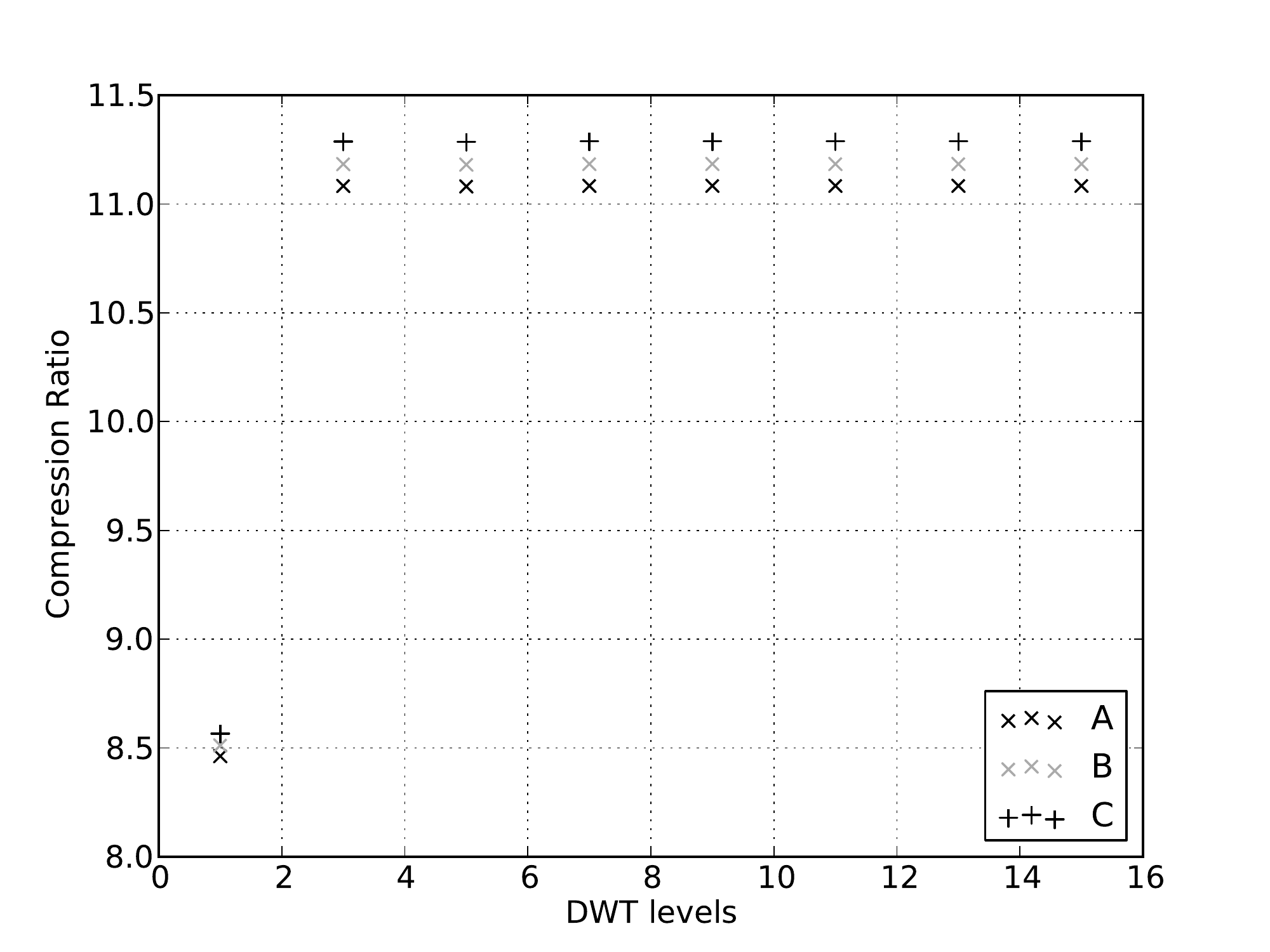}
  \caption{Compression Ratio vs DWT levels for datasets $A$, $B$
    and $C$.}
  \label{fig:levels-compress}
\end{figure}

\begin{figure}[ht]
  \includegraphics[width=\columnwidth]{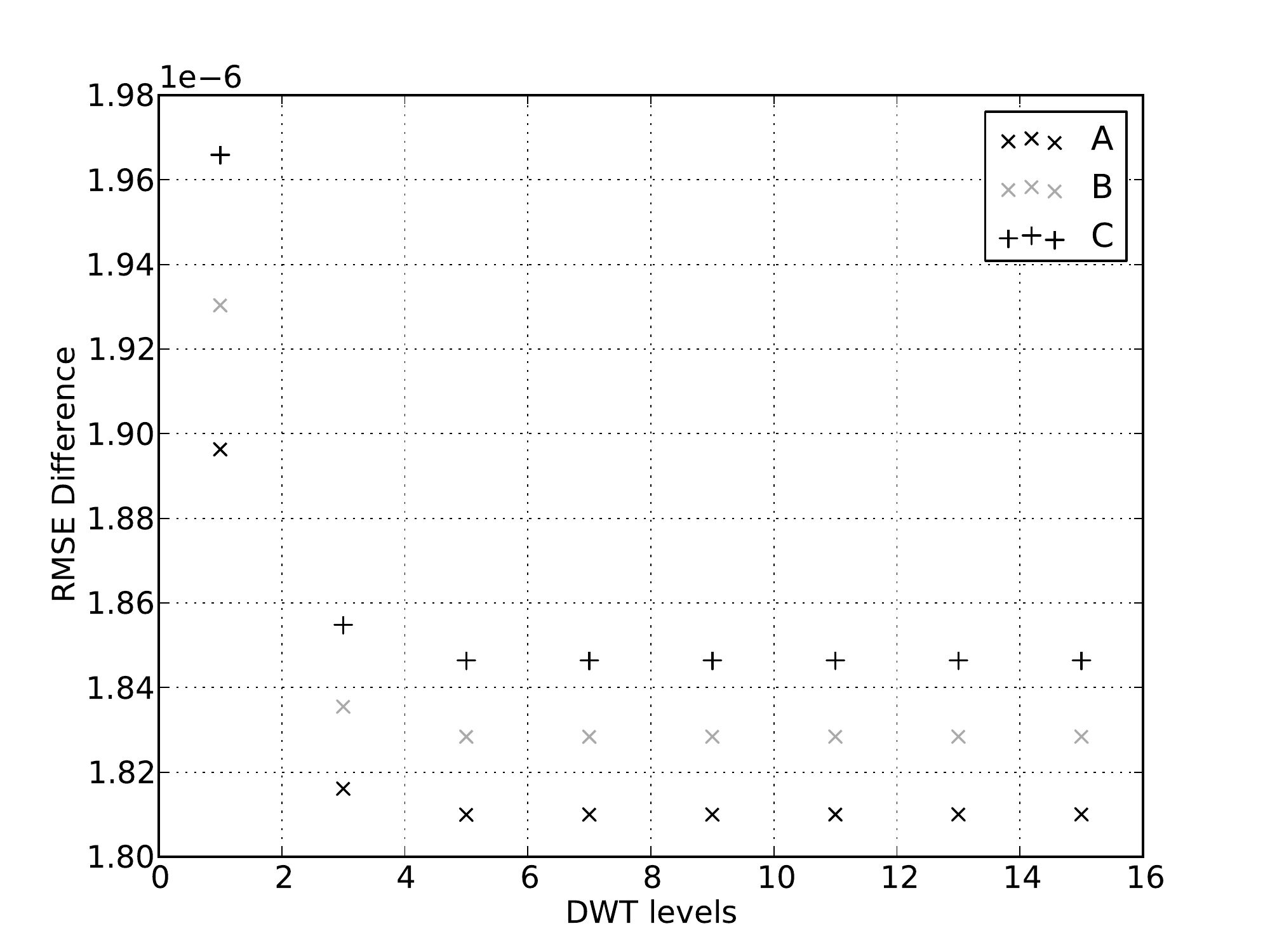}
  \caption{RMSE Difference vs DWT levels for datasets $A$, $B$ and $C$.}
  \label{fig:levels-rmse}
\end{figure}

Figures~\ref{fig:levels-compress} and~\ref{fig:levels-rmse} show that the number of
DWT levels did not significantly impact neither the compression ratio during the lossy 
compression or the RMSE difference, except for DWT=1.

\subsection{Completeness and Soundness} \label{ssec:cs}

\subsubsection{Completeness} \label{sssec:c}

\begin{figure}[ht]
  \includegraphics[width=\columnwidth]{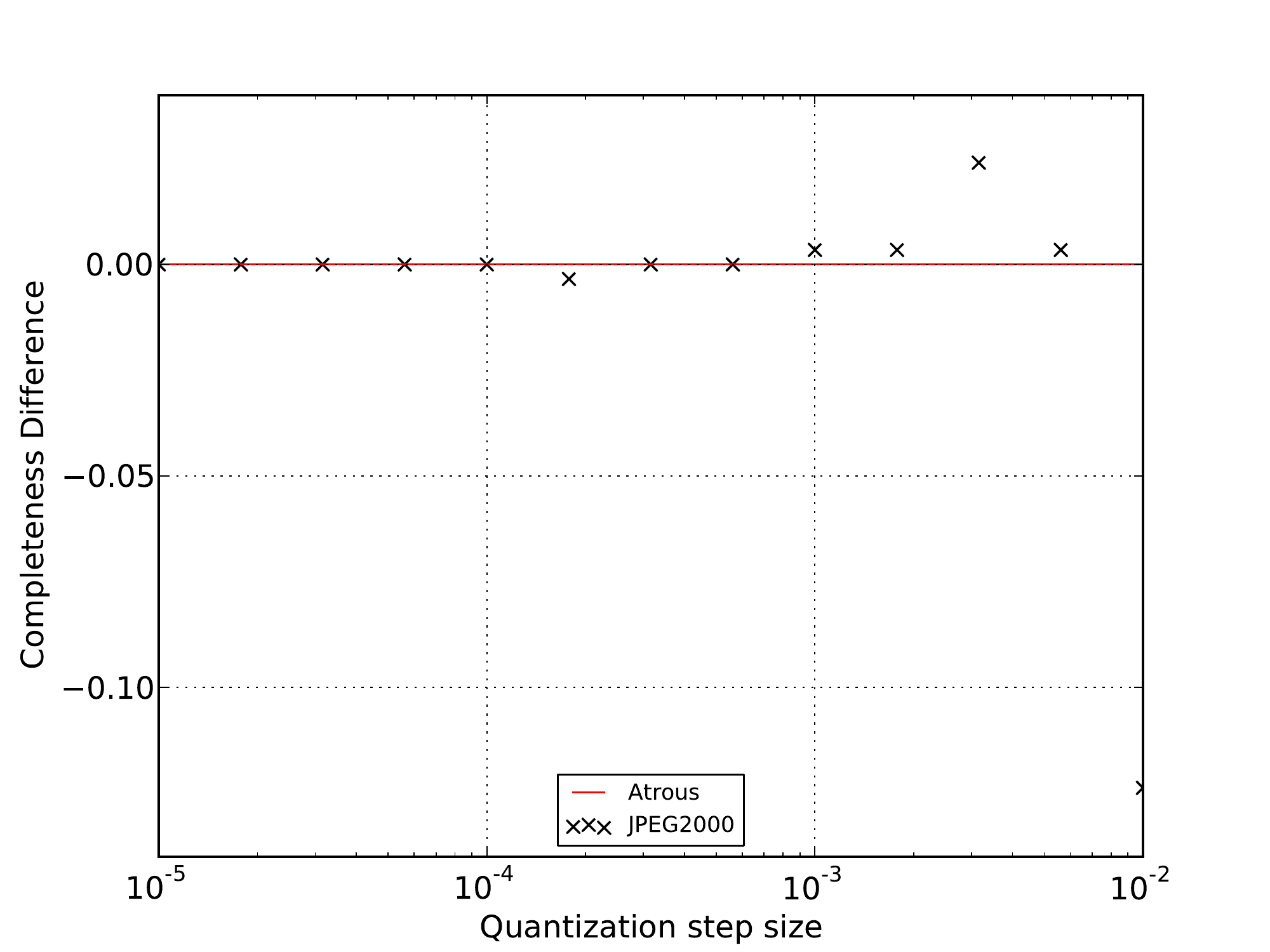}
  \caption{Completeness Difference vs Quantisation step size for dataset $A$.}
  \label{fig:A-qstep-comp}
\end{figure} 

In Figure~\ref{fig:A-qstep-comp} it can be seen that in dataset $A$ the source identification algorithm achieves equivalent
completeness when compressed for almost all values of quantisation step size.
There is a small dip just before an increase by
as much as 3\% in the
completeness of the sources identified. This peak of completeness at a
quantisation step size of $3\times10^{-3}$ corresponds with an extremely
high compression ratio of over $5\times10^2$. The final data point captured shows a
significant drop in completeness which simply corresponds to the image
eventually losing all scientific quality at extremely high compression.

\begin{figure}[ht]
  \includegraphics[width=\columnwidth]{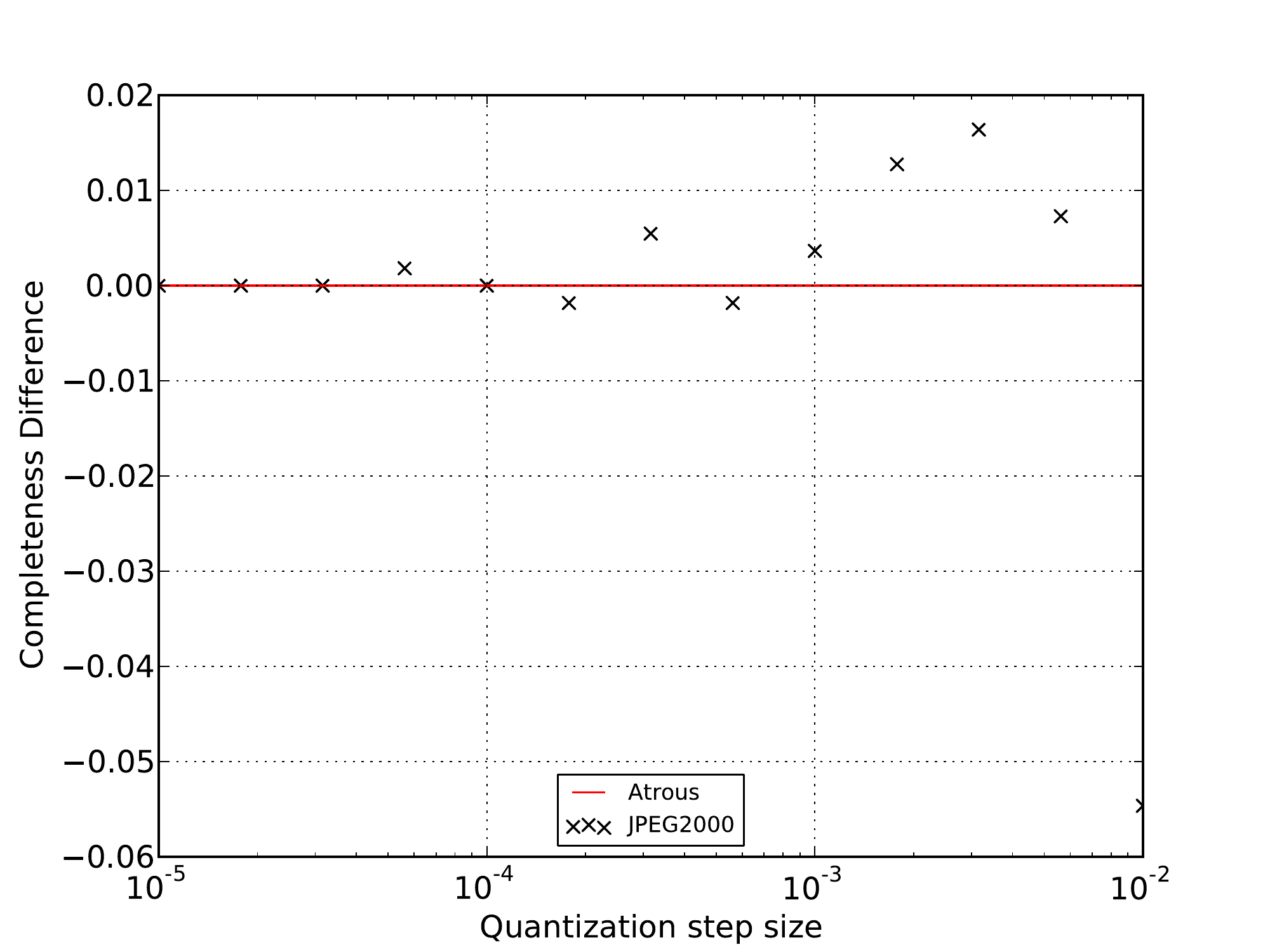}
  \caption{Completeness Difference vs Quantisation step size for dataset $B$.}
  \label{fig:B-qstep-comp}
\end{figure}

Datasets $B$ and $C$ show similar results. In Figure~\ref{fig:B-qstep-comp} we 
observe that \emph{Duchamp} achieves higher or equal completeness on a compressed dataset
$B$ for almost all data points. The completeness achieved in dataset $B$, was
almost as high as 2\%, occurring again at a high compression ratio of over
$4\times10^2$.

\begin{figure}[ht]
  \includegraphics[width=\columnwidth]{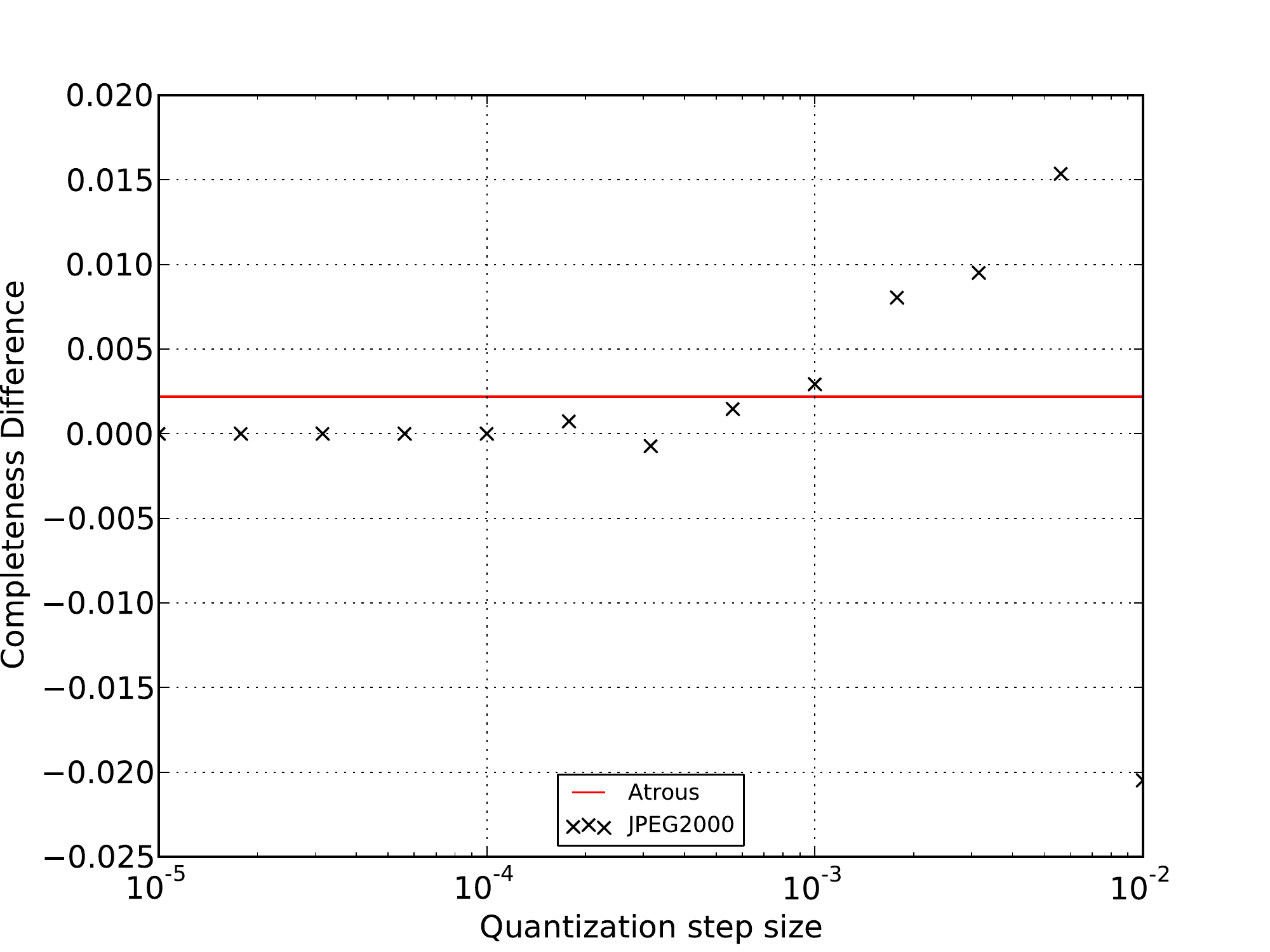}
  \caption{Completeness Difference vs Quantisation step size for dataset $C$.}
  \label{fig:C-qstep-comp}
\end{figure} 

Finally Figure~\ref{fig:C-qstep-comp}
shows a steady increase in the completeness of \emph{Duchamp} with respect to
the quantisation step size in dataset $C$. Dataset $C$ is particularly interesting 
as this dataset included, by far, the most sources. The vast majority of these sources
are fainter and thus more difficult to observe. This experiment clearly shows the
potential for JPEG2000 lossy compression to act as a denoising tool on spectral
datasets to achieve higher completeness in source identification.

\begin{figure}[ht]
  \includegraphics[width=\columnwidth]{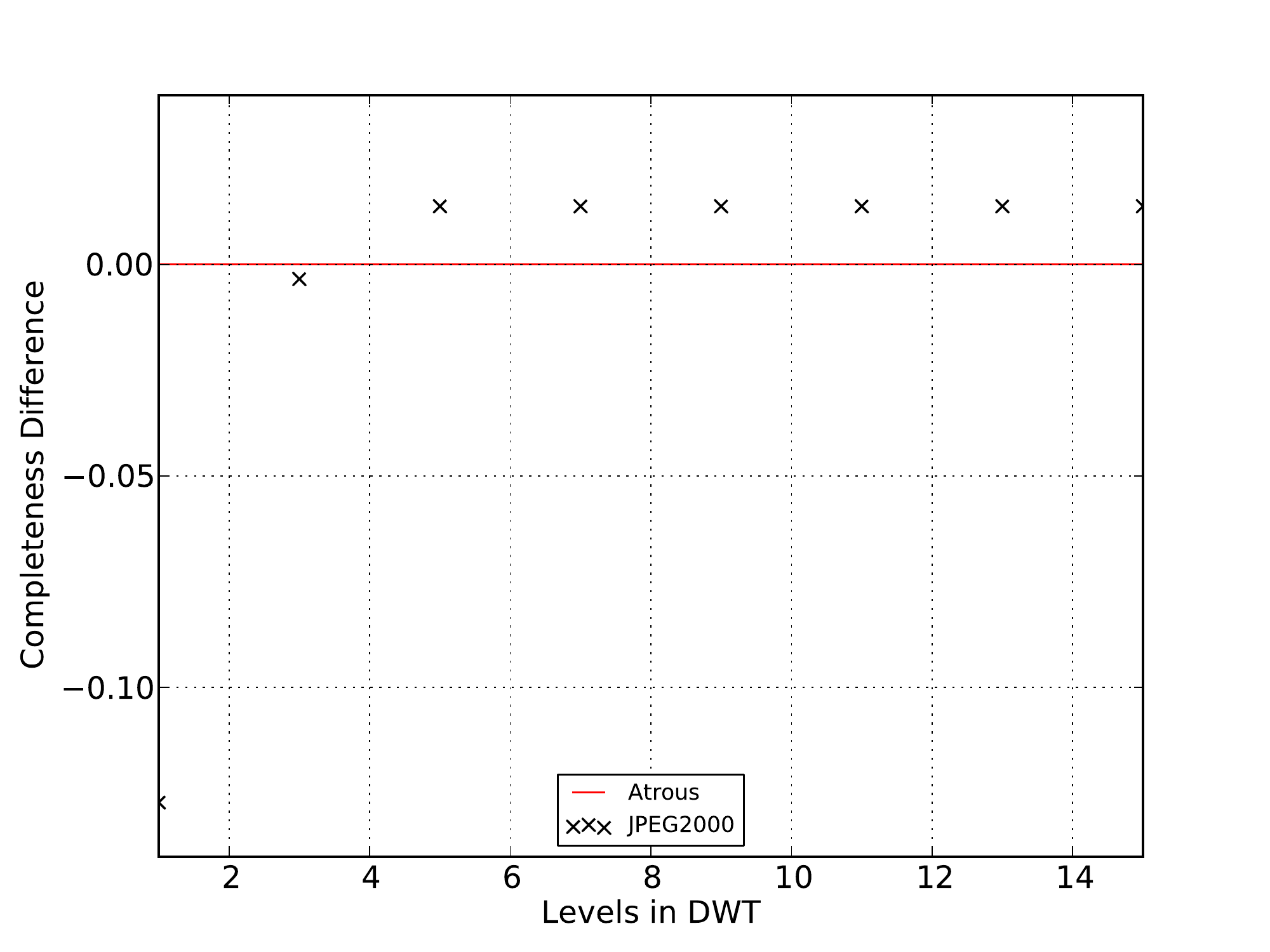}
  \caption{Completeness Difference vs DWT levels for dataset $A$.}
  \label{fig:A-levels-comp}
\end{figure}

\begin{figure}[ht]
  \includegraphics[width=\columnwidth]{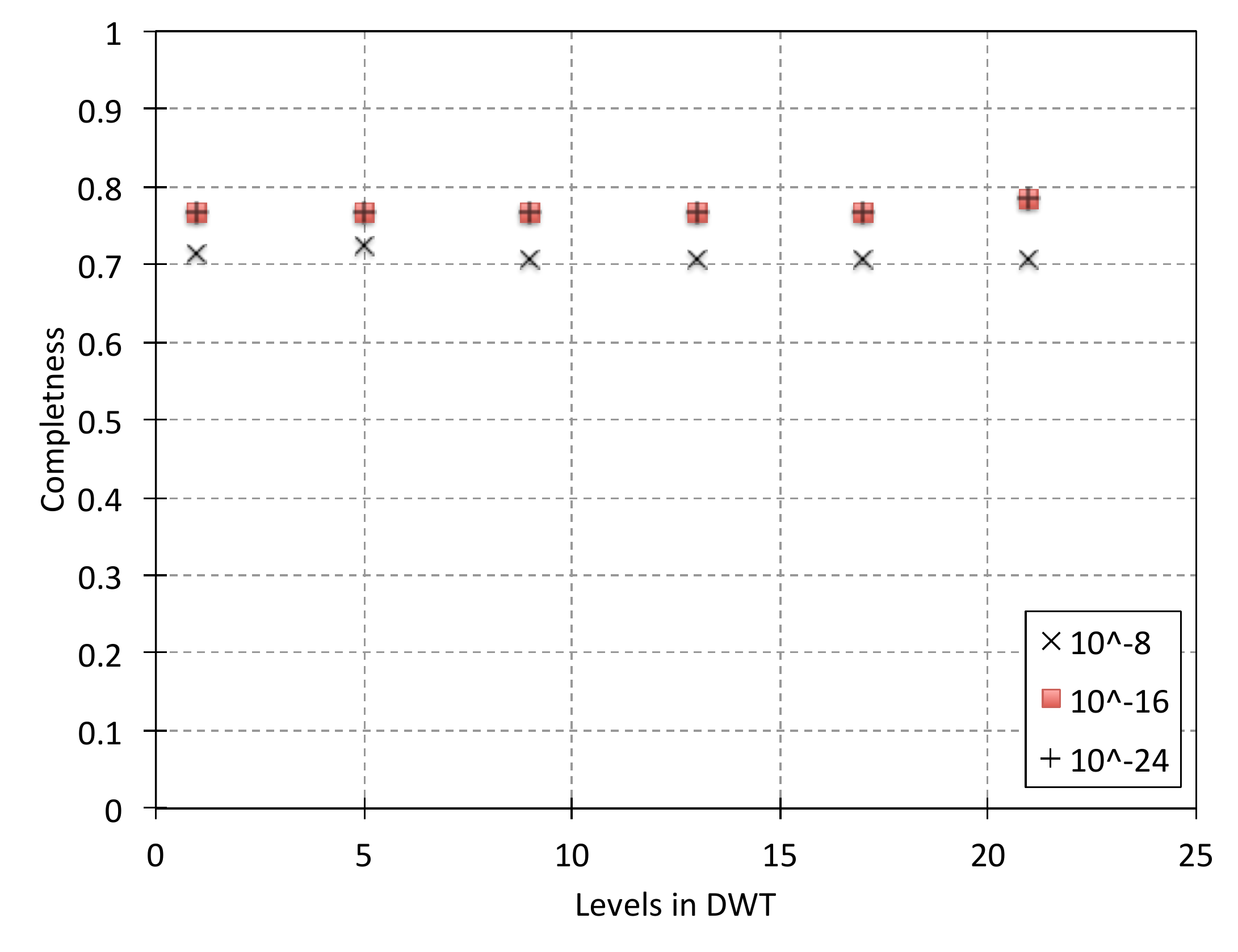}
  \caption{Completeness vs DWT levels for the quantisation step sizes $10^{-8}$, $10^{-16}$, $10^{-24}$ for all datasets.}
  \label{fig:DWT-comp}
\end{figure}

Figure~\ref{fig:A-levels-comp} obtained on dataset A shows that the number of DWT levels rather insignificantly
influences the completeness of the source finding. Figure~\ref{fig:DWT-comp} depicting the completeness obtained 
in experiments with all datasets is also showing almost no effect of the number of DWT levels on the source finding.
This means that most of the information about the sources are contained in the first few DWT levels, and
further encoding is unnecessary.

The final two parameters, precinct size and block size, had no effect on the completeness or the
soundness. This is somewhat expected as neither parameter has any direct
effect on any of the lossy components within the JPEG2000 compression algorithm,
rather they directly effect the lossless components e.g. run length encoding.

\begin{figure}[ht]
  \includegraphics[width=\columnwidth]{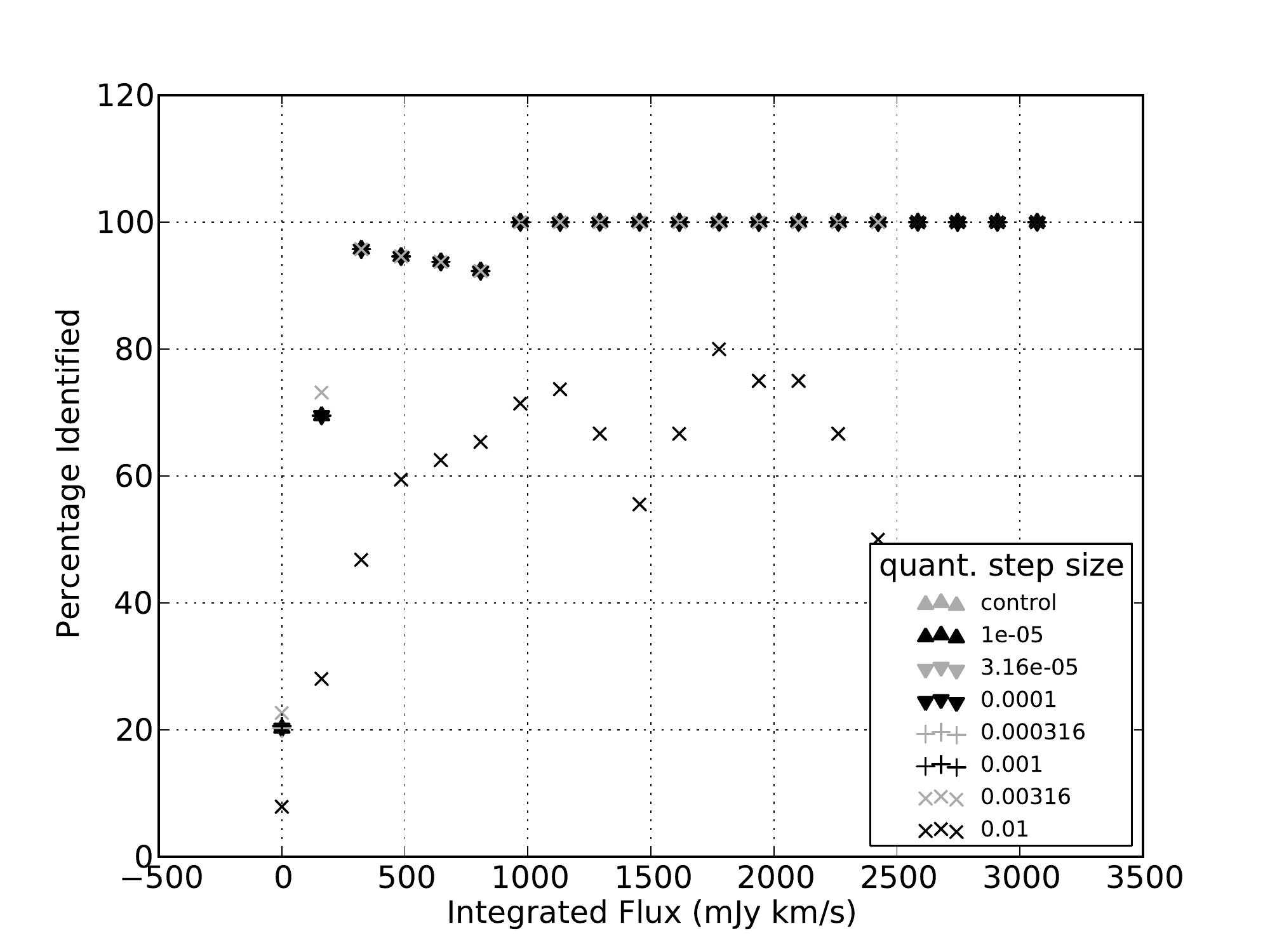}
  \caption{Completeness vs Integrated Flux for dataset $A$ compressed with a
    variety of different quantisation step sizes. The points for all the used quantisation step 
    sizes are superimposed, except for the step size $0.01$.}
  \label{fig:A-comp-flux}
\end{figure}

\begin{figure}[ht]
  \includegraphics[width=\columnwidth]{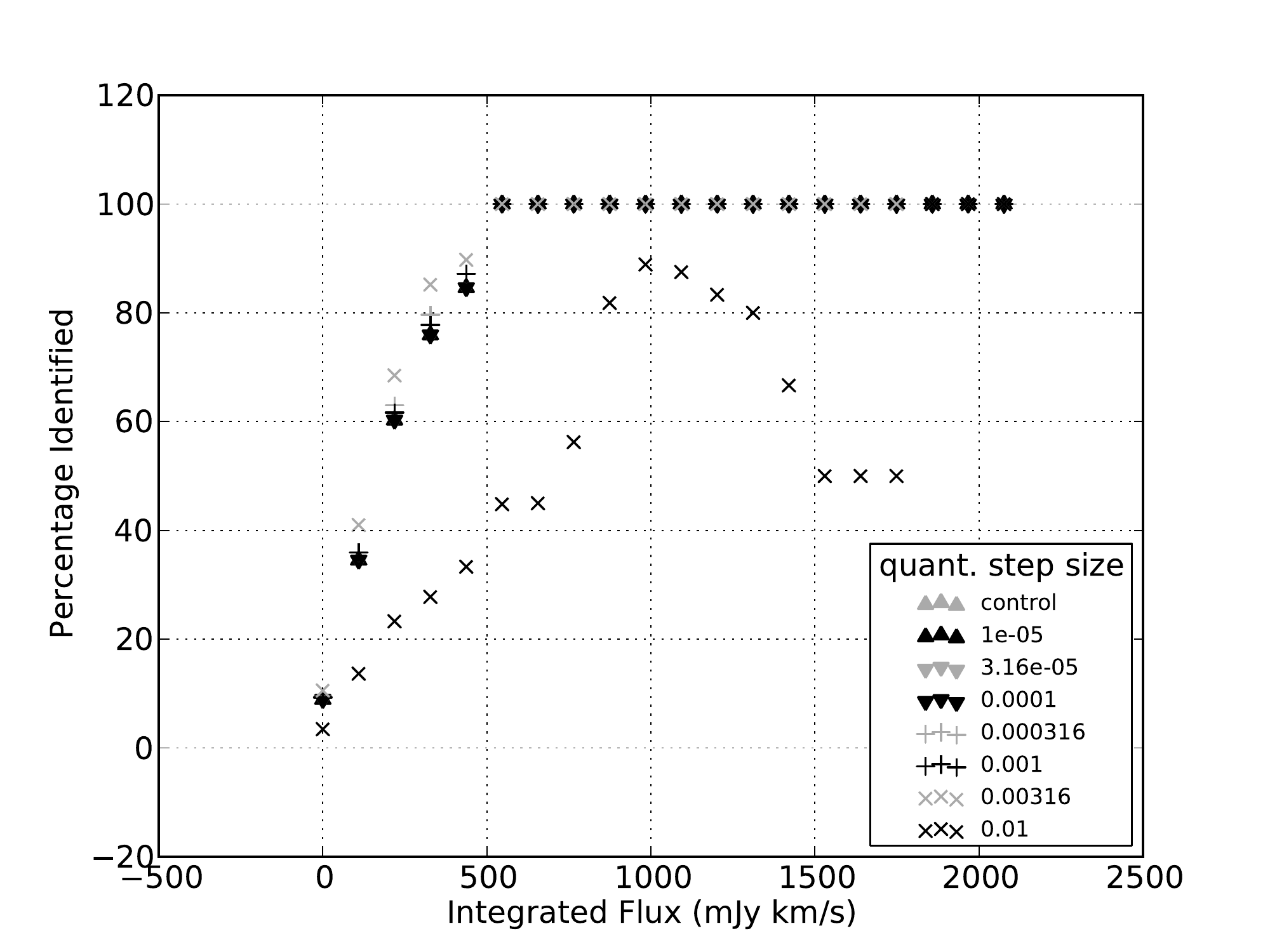}
  \caption{Completeness vs Integrated Flux for dataset $B$ compressed with a
    variety of different quantization step sizes. The points for all the used quantisation step 
    sizes are fully or partially superimposed, except for the step size $0.01$.}
  \label{fig:B-comp-flux}
\end{figure}

\begin{figure}[ht]
  \includegraphics[width=\columnwidth]{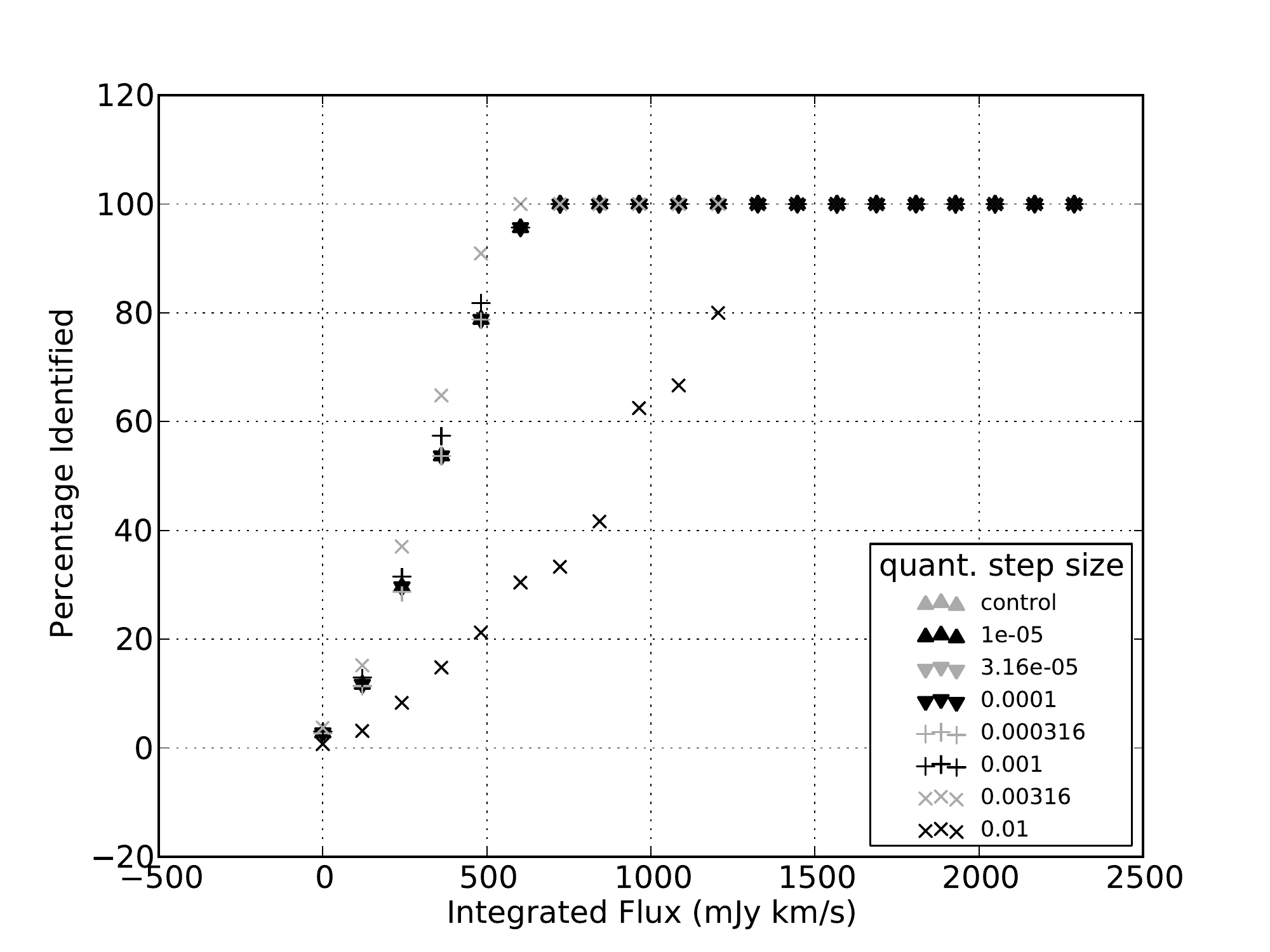}
  \caption{Completeness vs Integrated Flux for dataset $C$ compressed with a
    variety of different quantisation step sizes. The points for all the used quantisation step 
    sizes are fully or partially superimposed, except for the step size $0.01$.}
  \label{fig:C-comp-flux}
\end{figure}

Figures~\ref{fig:A-comp-flux},~\ref{fig:B-comp-flux} and~\ref{fig:C-comp-flux}
 show the completeness with respect to the
integrated flux of a source for each dataset. The clear upward trend in all graphs is due to the
fact that sources with higher total flux are easier
identified. The dark `x' data points correspond to the highest quantisation step size
where the image is extremely compressed, resulting in significant loss in
completeness. Finally across all datasets it is apparent that more low integrated
flux sources of less than 800 mJy per km/s are identified at higher quantisation step size
compressions, which again lends to the conclusion that JPEG2000 has a denoising effect 
increasing the signal to noise ratio and allowing previously undetectable by the \emph{Duchamp} 
sources to be identified.

\begin{table}[ht]
\centering
\begin{tabular}
{|p{0.13\columnwidth}|p{0.33\columnwidth}|p{0.33\columnwidth}|}
\hline
Dataset & Completeness & Soundness \\
\hline
A & 0.203 & 0.766 \\
B & 0.089 & 0.576 \\
C & 0.028 & 0.704 \\
\hline
\end{tabular}
\caption{The completeness and soundness of Duchamp on the uncompressed datasets
(Control).}
\label{tab:cs}
\end{table}

In dataset A we found an improvement from the original
completeness of 0.203 by $\sim$3\% to 0.23, in dataset B the improvement increased
the completeness from 0.089 to 0.11 and in dataset C the improvement had a peak
increase of completeness from 0.028 to 0.043. This result conclusively shows
JPEG2000 to be having a denoising effect on the simulated DINGO cube dataset.

\subsubsection{Soundness} \label{sssec:s}

Completeness cannot be considered independently from the soundness of
the source identification algorithms.

\begin{figure}[ht]
  \includegraphics[width=\columnwidth]{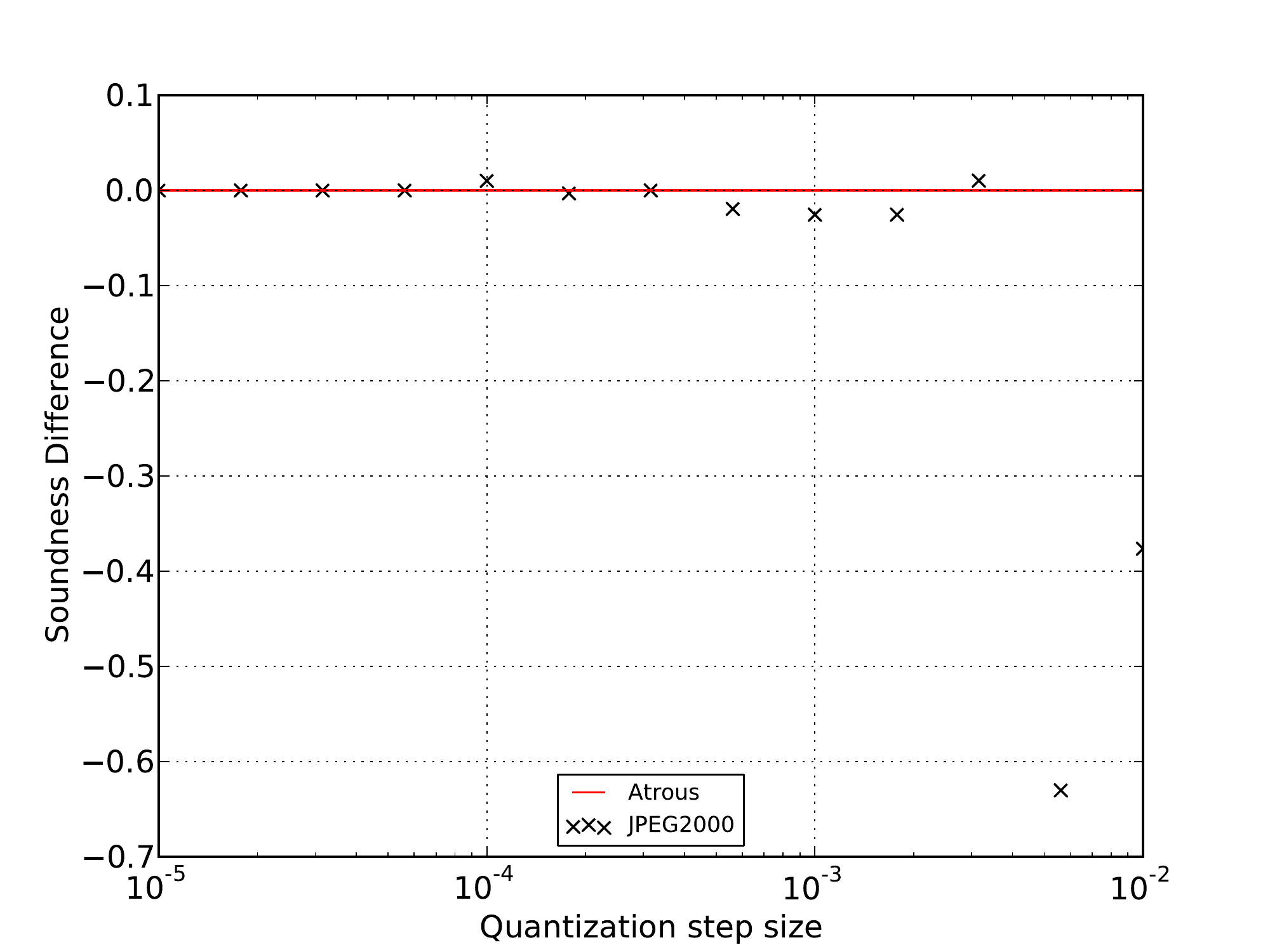}
  \caption{Soundness Difference vs Quantisation step size for dataset $A$.}
  \label{fig:A-qstep-sound}
\end{figure} 

\begin{figure}[ht]
  \includegraphics[width=\columnwidth]{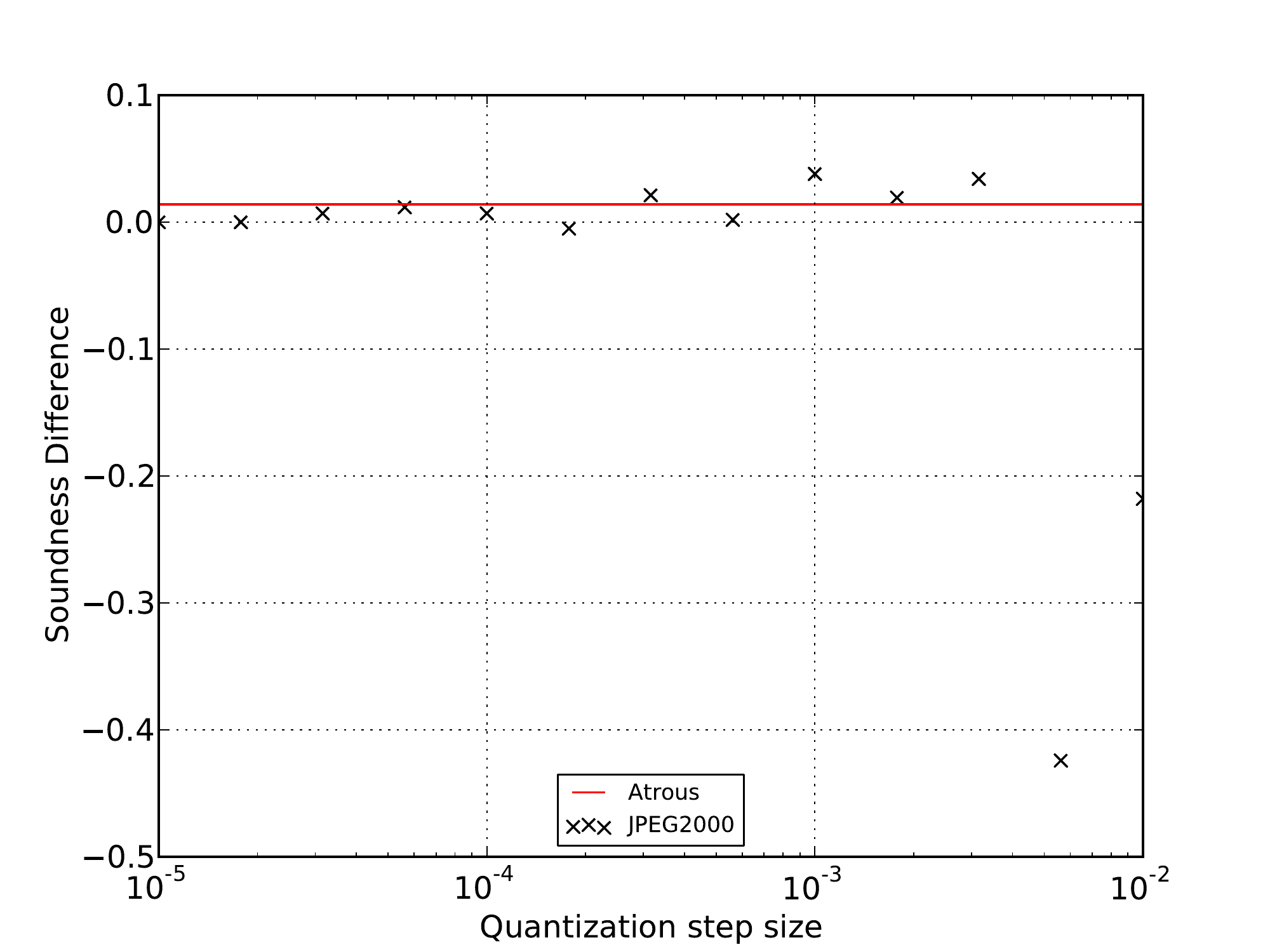}
  \caption{Soundness Difference vs Quantisation step size for dataset $B$.}
  \label{fig:B-qstep-sound}
\end{figure}

\begin{figure}[ht]
  \includegraphics[width=\columnwidth]{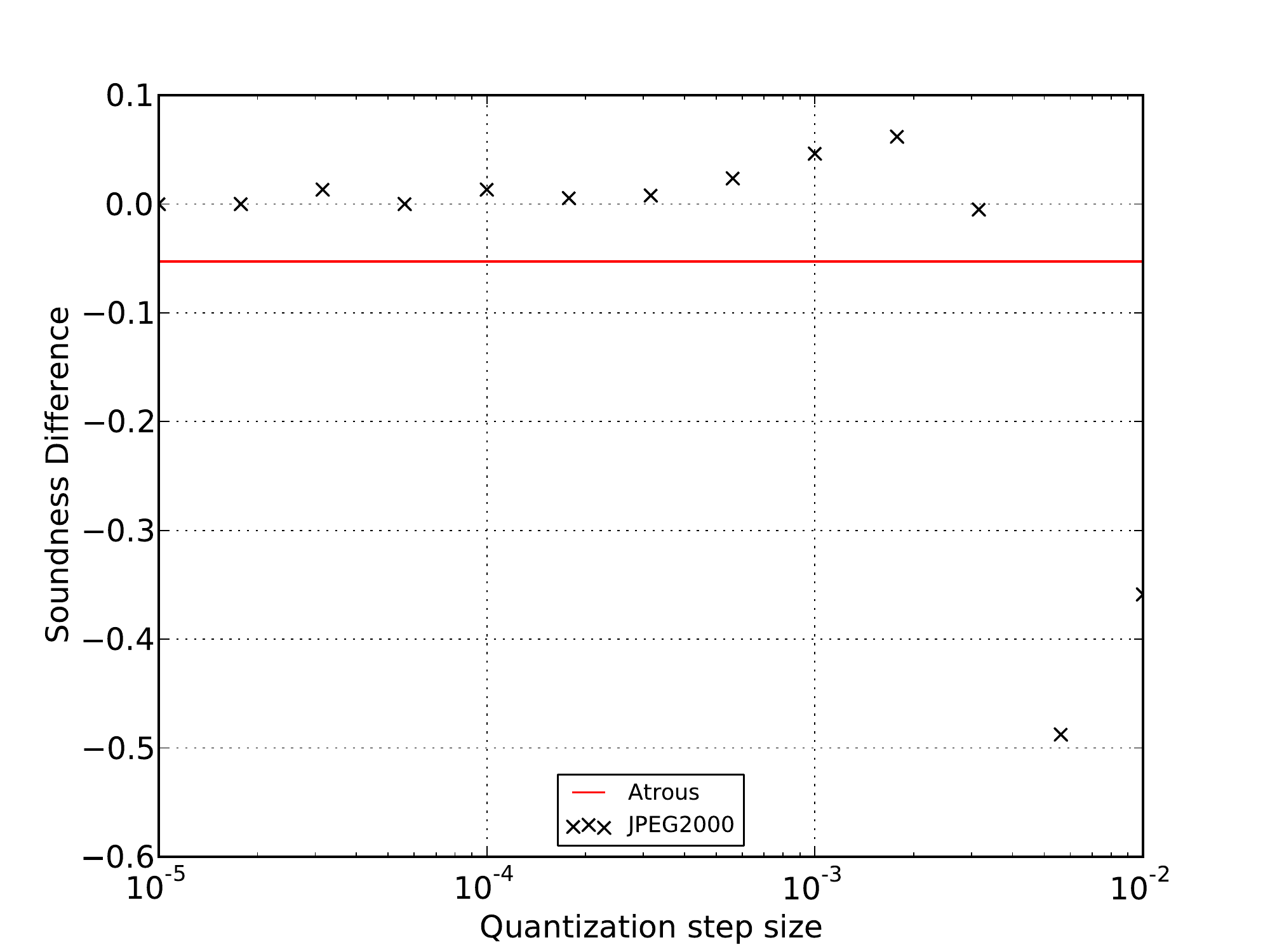}
  \caption{Soundness Difference vs Quantisation step size for dataset $C$.}
  \label{fig:C-qstep-sound}
\end{figure} 

\begin{figure}[ht]
  \includegraphics[width=\columnwidth]{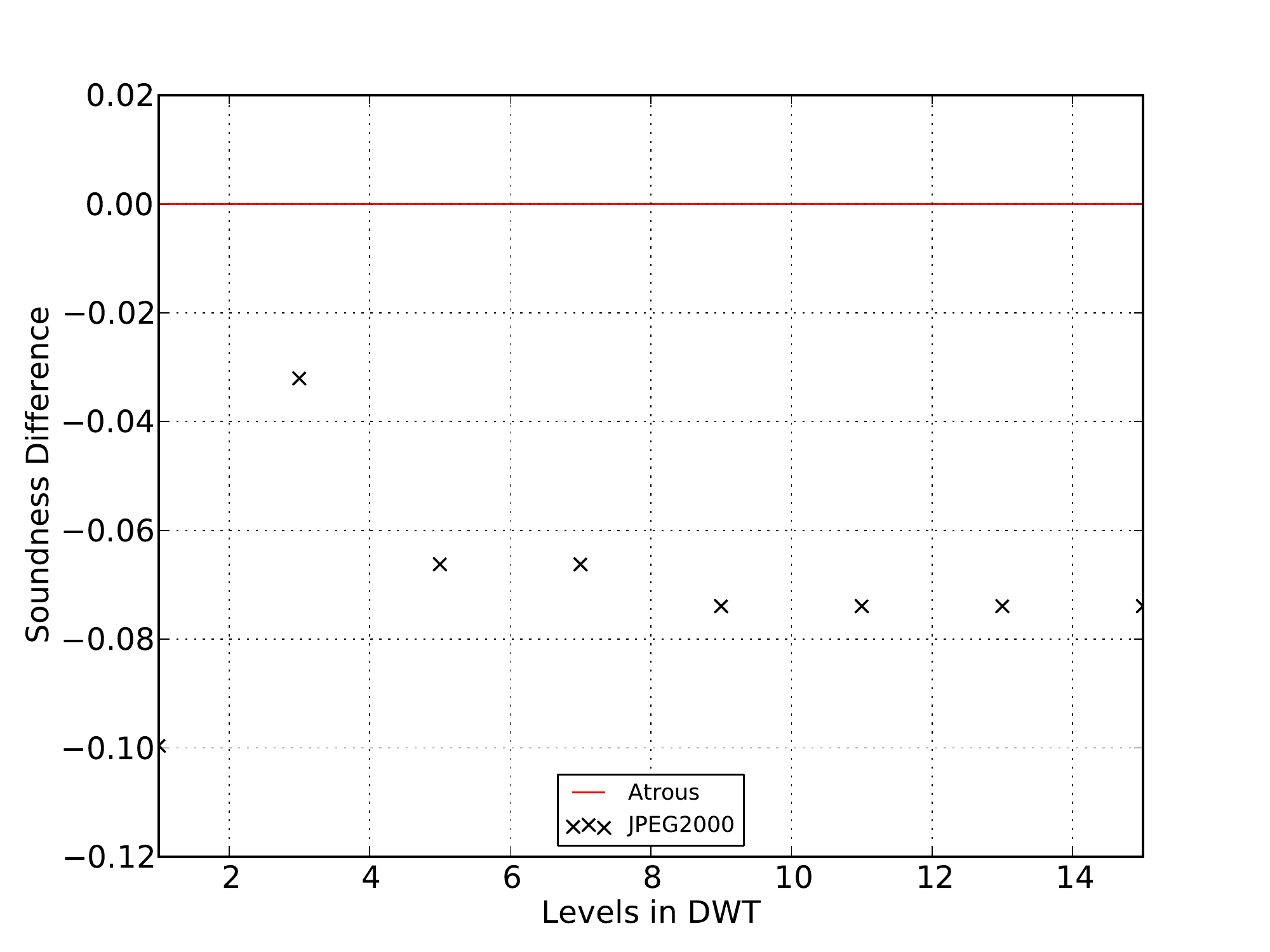}
  \caption{Soundness Difference vs DWT levels for dataset $A$.}
  \label{fig:A-levels-sound}
\end{figure}

Figure~\ref{fig:A-qstep-sound} shows the comparison of the soundness on dataset $A$ for the  \emph{Duchamp}'s ``\emph{algorithme \`{a} trous}'' (red line) and JPEG2000. One can see that the soundness remains fairly
consistent for JPEG2000 until it drops off at the quantisation step approximately $4\times10^{-2}$. A small dip does exist in soundness at a quantization step size
of approximately $1\times10^{-3}$. If we reference the respective completeness graph,
Figure~\ref{fig:A-qstep-comp},
we notice that the completeness appears to increase after this
dip. This result is not unexpected as when the compression
algorithm acts as a ``filter'' to the noise, the large structure instrumental noise will
not be lost through the lossy compression. These stronger pieces of noise
along with previously unidentified true sources are now more likely to be
identified as a source. In fact with any source identification method that has a less than perfect
soundness, if the completeness increases and the soundness remains stable both
the true positives and false positives will increase.                                      
It is therefore quite probable that if just short of enough denoising is performed to
identify a new true positive, there may be new false positives in our             
catalogue.  

The soundness of \emph{Duchamp} on dataset $B$ and $C$ stays the same or increases under 
JPEG2000 lossy compression for all data points excluding the final two
as observed in Figures~\ref{fig:B-qstep-sound} and~\ref{fig:C-qstep-sound}. In
particular the soundness was only ever increased or exactly the same, at the 
average quantisation step size where completeness peaked.
At high compression ratios JPEG2000 is known to
occasionally cause ringing artefacts~\citep{ringing}. If these ringing artefacts
were to occur widely across the image the source finder may identify them as
sources. This could potentially explain why we observe a loss in soundness at
the highest compression ratios.
We can note however, that this loss in soundness occurs after the average
quantisation step size found to give peak completeness improvement to source
identification. We therefore still find lossy JPEG2000 compression to improve
completeness without loss in soundness at appropriate quantisation step sizes.

For all datasets however, the soundness fell dramatically for the final two 
quantisation step size. We can thus conclude that a quantisation step size of
higher than $6\times10^{-3}$ will negatively effect accurate source
identification.

The number of DWT levels on the other hand, had a surprisingly negative correlation with the
soundness of the \emph{Duchamp} source finder as seen in
Figure~\ref{fig:A-levels-sound}. It should, however, be noted that
all data points in this Figure indicated the soundness to decrease after
compression. As we have seen this is not the case, when exploring over the
quantisation step size space, we can conclude that this is the result of high default 
quantisation step size of $1/256$ used.

Overall we identify the quantisation step size to be the dominating parameter
where a value of between $1\times10^{-3}$ and $3\times10^{-3}$
was found to be optimal to improve the completeness and soundness of source identification.
Across all datasets the completeness and soundness trended upwards with respect
to quantisation step size up to and including these data points. To optimise on
particular datasets we would recommend searching between these values.

\subsubsection{The ``\emph{algorithme \`{a} trous}''}
\label{sssec:atrous}

\begin{table}[ht]
\centering
\begin{tabular} {|l|l|l|}
\hline
Dataset & \emph{Duchamp} with atrous & \emph{Duchamp} with JP2 \\
\hline
A & 292min 32s & 24min 12s \\
B & 284min 16s & 22min 43s \\
C & 315min 11s & 27min 6s \\
\hline
\end{tabular}
\caption{The time taken for \emph{Duchamp} with the wavelet reconstruction,
versus time taken for \emph{Duchamp} with encoding and decoding with JPEG2000}
\label{tab:atrous_v_jp2_time}
\end{table}

Overall the denoising effect of wavelet reconstruction in \emph{Duchamp} was outperformed
by the denoising effect of JPEG2000 image compression. In fact \emph{Duchamp's} wavelet reconstruction had little
effect at all on completeness and soundness. As observed in Figure~\ref{fig:C-qstep-sound} dataset $C$ was the only dataset to see an increase in
source identification completeness after reconstruction, where that increase was
only 0.25\%. The soundness was
positively effected in datasets $B$ in Figure~\ref{fig:B-qstep-sound}, but
notably less so than the peak soundnesses found by simply compressing the image
cube. Dataset $A$ saw no effect on
either soundness or completeness after a \emph{Duchamp} wavelet reconstruction. 

This lack of significant effect can
be attributed to the parameter space of the wavelet reconstruction not being 
fully explored. It is also important to note that the
\emph{Duchamp} source finder's wavelet reconstruction has been improved in Duchamp's
successor Selavy, to the reconstruction used by the 2D-1D wavelet reconstruction
source identification algorithm~\citep{2d1d}. 
The ``\emph{algorithme \`{a} trous}'' found in \emph{Duchamp} was also far more 
computationally expensive than the JPEG2000 compression algorithm. 
Table~\ref{tab:atrous_v_jp2_time} shows for each dataset how much longer
\emph{Duchamp's} wavelet reconstruction took in comparison to how long encoding
to JPEG2000, decoding back to FITS
and then executing \emph{Duchamp} without the wavelet reconstruction took.

It is, in fact, because of the computationally expensive nature of the wavelet
reconstruction algorithm that a larger set of parameters could not be explored.
Which lends to the hypothesis that the DWT in JPEG2000 as a denoising tool may
be preferred over the ``\emph{algorithme \`{a} trous}'' because while the DWT may
include the undesirable trait of shift variance, the ``\emph{algorithme \`{a} trous}''
is simply too slow for extremely large datasets. Furthermore, JPEG2000 can be
used to generate compressed preview datasets for quality control and visual exploration
of data. Once generated with optimal parameters the previews can be used for source 
finding purposes removing a need for prior denoising, and substantially improving
I/O performance due to the smaller size of compressed datasets.

\subsection{Source Parameterisation} \label{ssec:source_param}

\subsubsection{Right Ascension and Declination}

\begin{figure}[ht]
  \includegraphics[width=\columnwidth]{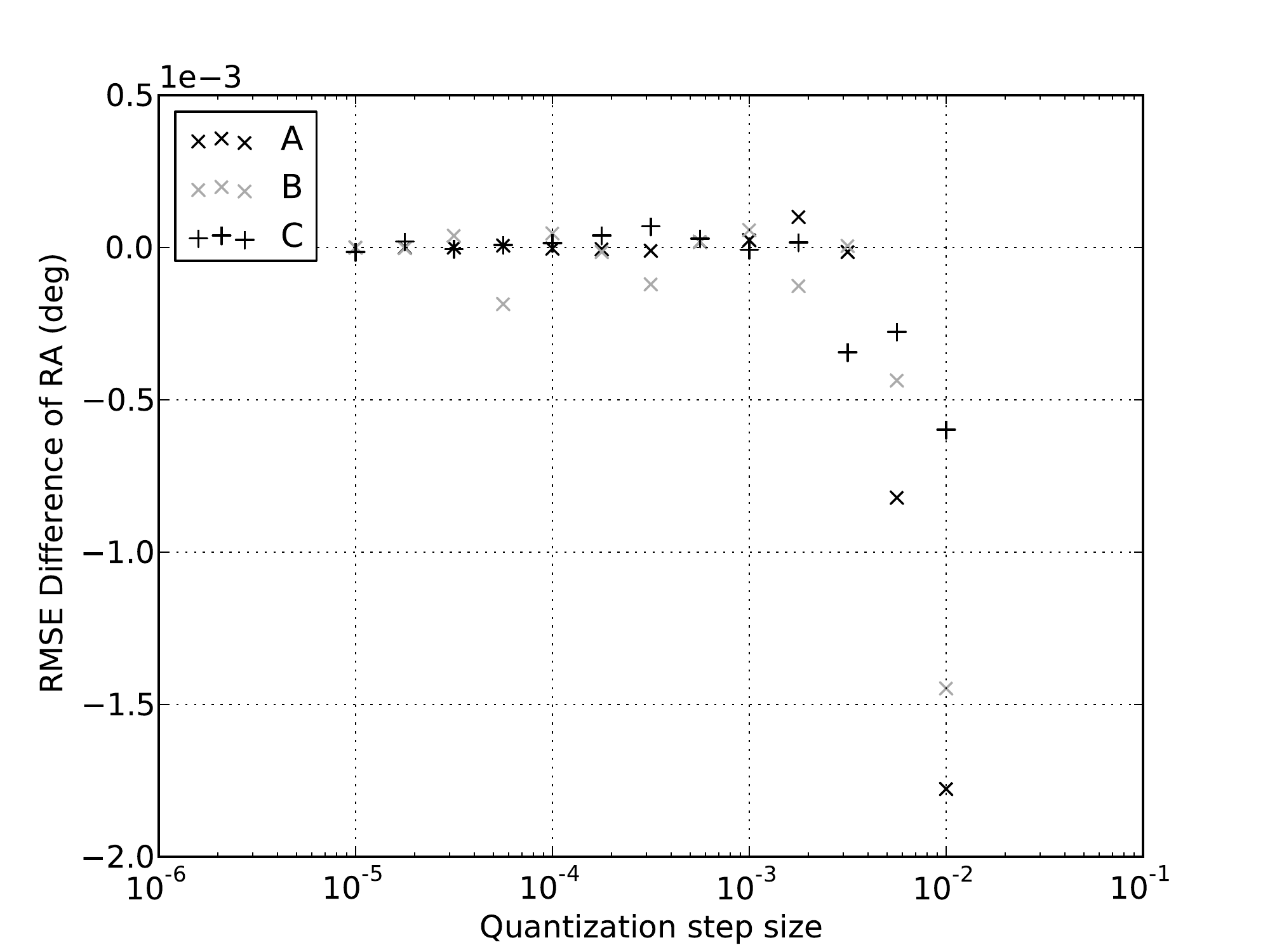}
  \caption{RMSE difference of RA vs Quantisation step size for all datasets.}
  \label{fig:qstep-rmse-ra}
\end{figure} 

\begin{figure}[ht]
  \includegraphics[width=\columnwidth]{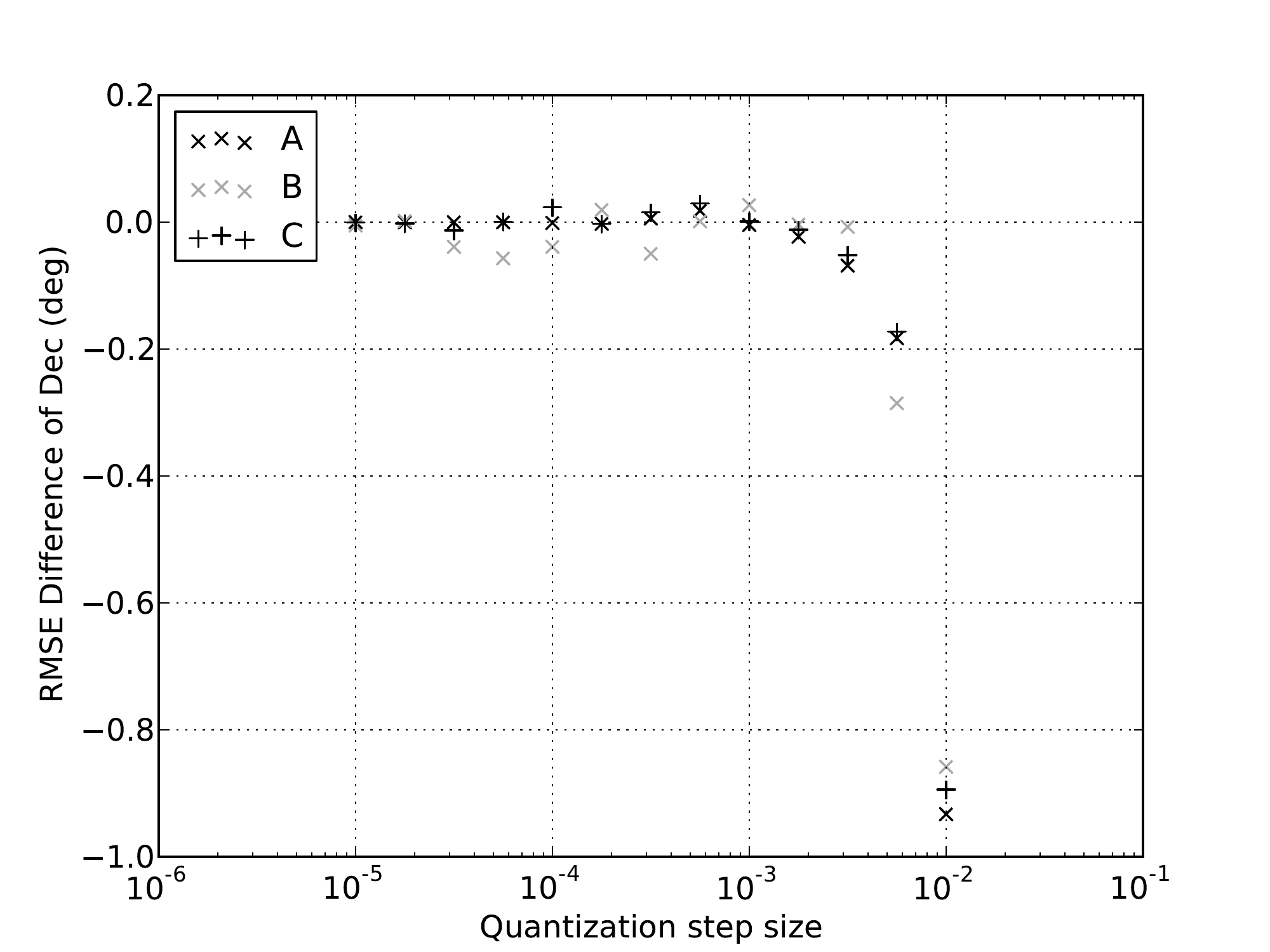}
  \caption{RMSE difference of Dec vs Quantisation step size for all
datasets.}
  \label{fig:qstep-rmse-dec}
\end{figure}

\begin{figure}[ht]
  \includegraphics[width=\columnwidth]{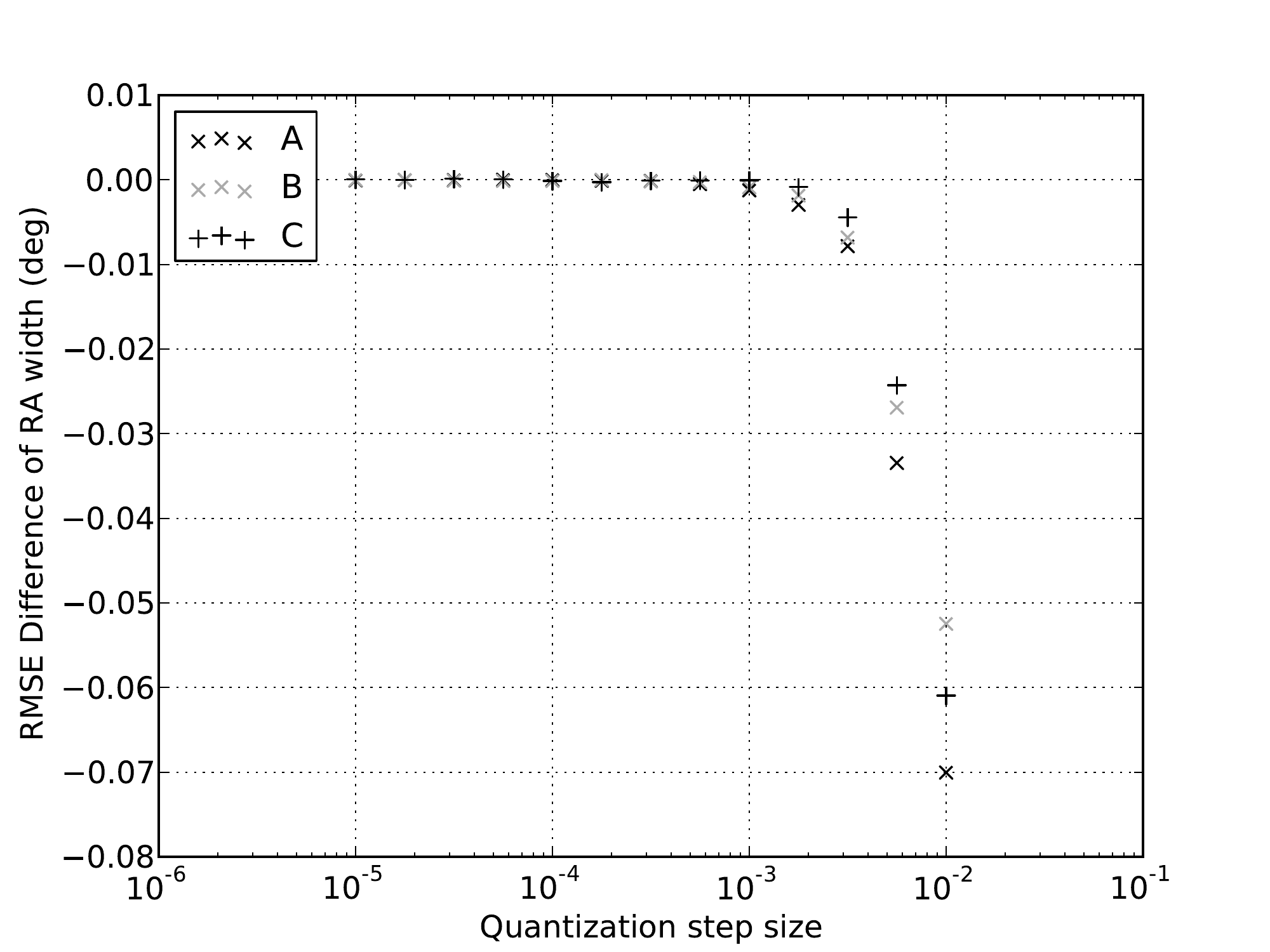}
  \caption{RMSE difference of RA width vs Quantisation step size for all
datasets.}
  \label{fig:qstep-rmse-wra}
\end{figure} 

\begin{figure}[ht]
  \includegraphics[width=\columnwidth]{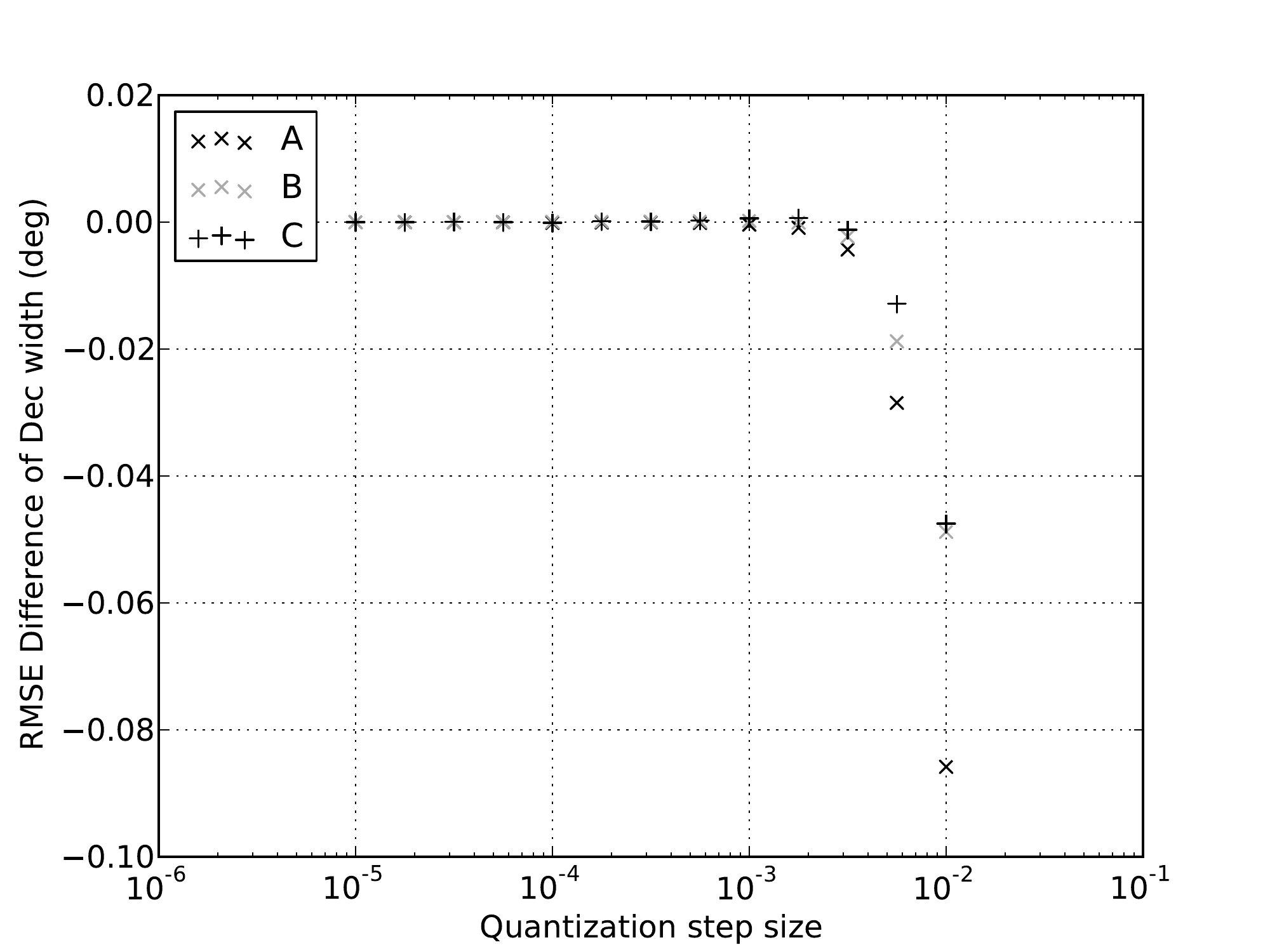}
  \caption{RMSE difference of Dec width vs Quantisation step size for all
datasets.}
  \label{fig:qstep-rmse-wdec}
\end{figure} 

The spatial coordinates of sources identified by \emph{Duchamp} appear to be
more negatively affected by JPEG2000's lossy compression. We can observe
in Figure~\ref{fig:qstep-rmse-ra} that when
compressed with a quantisation step size of higher than approximately 
$3\times10^{-4}$, the accuracy of RA source parameterisation drops off. This is
notably well before the quantisation step size resulting in a peak in completeness.

While the Dec source parameterisation in Figure~\ref{fig:qstep-rmse-dec} follows
a similar trend, we identified
what appeared to be an error in either the true-set catalogue's Dec parameter or
\emph{Duchamp's} Dec parameterisation. Sources that were identified with the
exact same centroid voxel as found in the true catalogue were often found to
have a Dec difference by as much as a degree. We can therefore not make any
conclusive judgement from our results with regard to Dec source parameterisation.

Figures~\ref{fig:qstep-rmse-wra} and~\ref{fig:qstep-rmse-wdec} both show a loss
in the accuracy of spatial width parameterisation at high compression ratios. As the
voxels become more correlated to each other through compression the soft edges
of the sources appear to be either stretched above or below the threshold of a
source. This directly shows that at high compression ratios of the JPEG2000
lossy compression algorithm, the scientific quality of radio astronomy data is
negatively effected. At much more reasonable compression ratios, however, the
effect is zero.

Overall source parameterisation of spatial width remained unaffected until
compressed with a quantisation step size greater than $1\times10^{-3}$ that 
corresponds to compression ratio approximately 1:100 (see Figure~\ref{fig:qstep-compress}). 

\subsubsection{Integrated Flux} \label{sssec:intflux}

\begin{figure}
  \centering
  \includegraphics[width=\columnwidth]{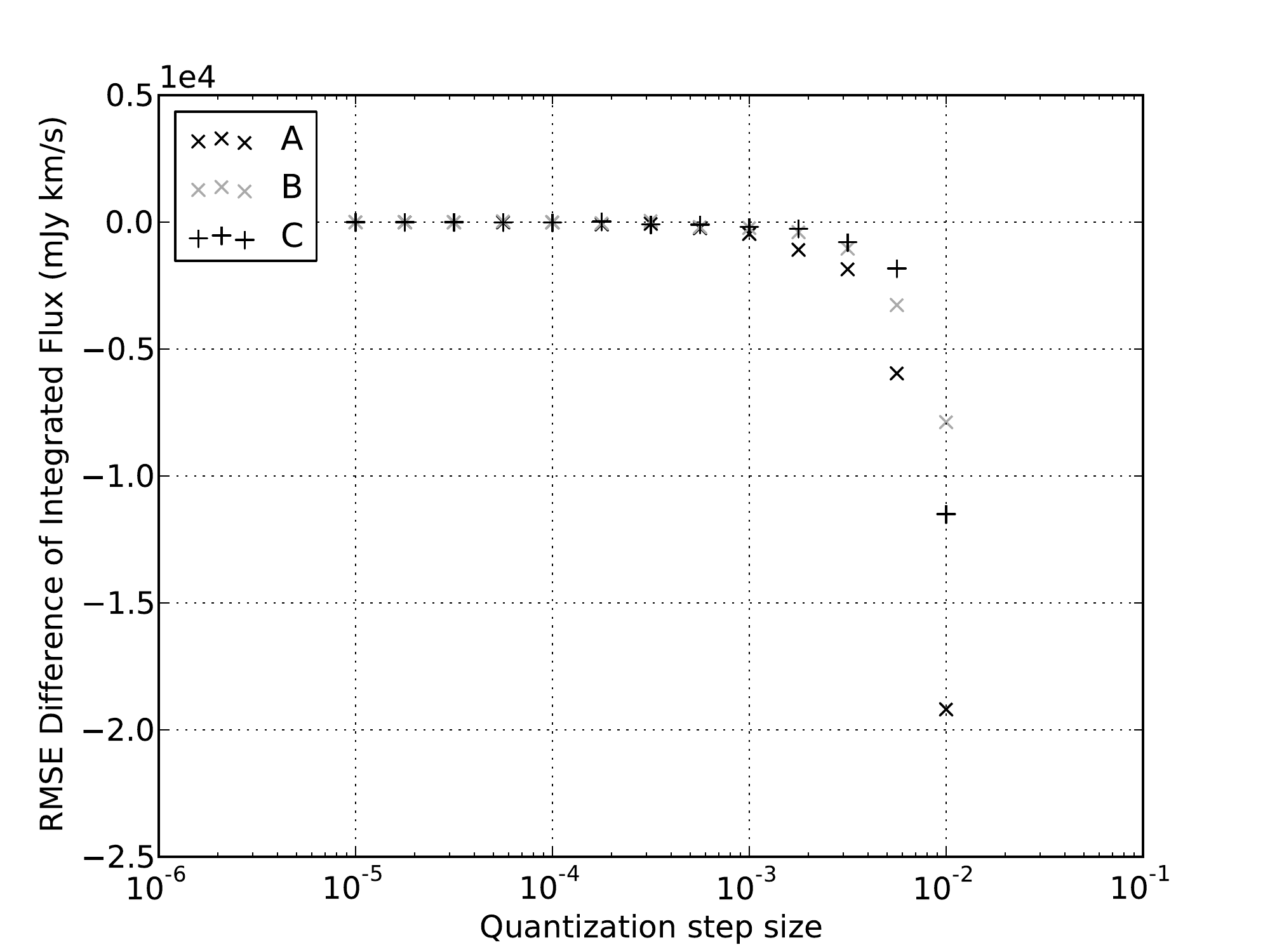}
  \caption{RMSE difference of Integrated Flux vs Quantisation step size for
all datasets.}
  \label{fig:qstep-rmse-intflux}
\end{figure}

Figure~\ref{fig:qstep-rmse-intflux} shows that across all datasets the
parameterisation of integrated flux performs poorly at high compression rates
(low quantisation step sizes). At high quantisation step sizes the domain of
wavelet coefficients becomes more discretised. As these coefficients become more
discretised the reconstruction will result in differing wavelet coefficient amplitudes. While
the wavelet transform will still be capable of maintaining the structure of
sources relative to the rest of the image, the original pixel amplitudes will
change after reconstruction. Thus while the sources may still have amplitudes
higher than neighbouring voxels, the actual value of the amplitude will have
changed and while completeness can be increased, integrated flux source
parameterisation is damaged.

In Section~\ref{ssec:cs} it was identified that the peak completeness and 
soundness occurred between quantisation step sizes of $1\times10^{-3}$
and $3\times10^{-3}$. We can clearly see in
Figure~\ref{fig:qstep-rmse-intflux} that by this point the source
parameterisation of the source identification algorithm had been damaged.

We thus conclude, that in order to maintain as accurate as possible source 
parameterisation while using JPEG2000's lossy compression algorithm, one should
not exceed the conservative limit of a compression ratio of 1:100 (or a quantisation
step of approximately $1\times10^{-3}$). Any kind of measurable negative impact
on source parametrisation had not been observed at all on the compression ratios
below 12, thus we can conclude that the compression did not affect the data in any
negative way.

\section{Conclusion} \label{cha:conc}

JPEG2000 has been found to have a negligible effect on the scientific quality of
radio astronomy imagery up to a compression ratio of approximately 12.
Thereafter, source parameterisation would progressively become less accurate.
At the same time, the completeness and soundness of source finding 
was increasing up to the compression ratios 1:100 and higher by as much
as 3\% and 7\% correspondently, as the result of noise filtering effect of the lossy 
wavelet JPEG2000 compression algorithm.

While the increase in completeness and soundness only occurred when source
parameterisation had become less accurate, the result may still be useful. Further
study is necessary to compare introduced errors due to the compression with other
errors already present in the data. Such a study needs to be done with real rather
than synthetic data.

JPEG2000 encoding had been found an order of 
magnitude faster than the ``\emph{algorithme \`{a} trous}'' used by the \emph{Duchamp}, 
and yet our results indicated the denoising effects of both to be comparable with 
JPEG2000 being slightly better. Future work should be
placed into the investigation and implementation of a DWT based wavelet
reconstruction, with quantisation and thresholding, to be used in source finders.

\section{Acknowledgements}
A special thanks Prof. Amitava Datta for his guidance while preparing this work.
Kakadu Software Ltd. and personally Prof. David Taubman for allowing us to freely use the software. 
Dr. Martin Meyer and Dr. Tobias Westmeier for their support in the use of the simulated DINGO 
cube dataset, and Dr. Attila Popping for other guidance in the Radio Astronomy domain.

\bibliography{references}
\end{document}